\documentclass[journal]{IEEEtran}

\usepackage[pdftex]{graphicx}
\graphicspath{{./figures/}{./figures/TFPatterns/}}
\DeclareGraphicsExtensions{.pdf,.jpeg,.png,.eps}

\ifCLASSINFOpdf
   \usepackage[pdftex]{graphicx}
   \graphicspath{{./figures/}}
   \DeclareGraphicsExtensions{.pdf,.jpeg,.png}
\else
\fi
\usepackage{amsmath}
\usepackage{mathtools}

\usepackage{algorithmic,algorithm}

\usepackage{array}
\usepackage{booktabs}

\usepackage{multicol,multirow,hhline}

\ifCLASSOPTIONcompsoc
  \usepackage[caption=false,font=normalsize,labelfont=sf,textfont=sf]{subfig}
\else
  \usepackage[caption=false,font=footnotesize]{subfig}
\fi
\usepackage{url}

\usepackage{pgf,pgfplotstable,tikz}
\usetikzlibrary{arrows,decorations.pathreplacing,calc}
\pgfplotsset{compat=1.10}
\definecolor{lred}{rgb}{0.6,0.2,0.}
\definecolor{lgray}{rgb}{0.6,0.6,0.6}
\usepackage[T1]{fontenc}

\usepackage{amssymb,amsmath,amsthm}
\usepackage{xifthen}
\usepackage{graphicx,xcolor}
\graphicspath{{figures/}}

\newcommand{\NLOp}[1]{\mathcal{#1}}
\newcommand{\ing}[1]{\mathsf{#1}}
\newcommand{\Rn}[1]{\ifthenelse{\equal{#1}{}}{\mathbb{R}}{\mathbb{R}^{\ing{#1}}}}
\newcommand{\Cn}[1]{\mathbb{C}^{\ing{#1}}}
\newcommand{\Rset}[2]{{~\in \Rn{\ing{#1} \times \ing{#2}}}}
\newcommand{\Cset}[2]{{~\in \Cn{\ing{#1} \times \ing{#2}}}}
\newcommand{\vect}[1]{\boldsymbol{\mathbf{\MakeLowercase{#1}}}}
\newcommand{\mtrx}[1]{\boldsymbol{\mathbf{\MakeUppercase{#1}}}}
\newcommand{\var}[1]{\mathrm{#1}}

\newcommand{\norm}[2]{\|#1\|_{#2}}

\newcommand{\htransp}[1]{#1^{\mathsf{H}}}

\newcommand{\ind}[2]{\ifthenelse{\equal{#2}{}}{\chi_{#1}}{\chi_{#1}\left( #2 \right)}}

\newcommand{\iter}[2]{#1^{\ing{(#2)}}}

\DeclareMathOperator*{\argmin}{argmin}
\DeclareMathOperator*{\argmax}{arg\,max\,}
\DeclareMathOperator*{\minim}{minimize\,}

\DeclareMathOperator*{\sign}{sgn\,}
\newcommand{\abs}[1]{\vert#1\vert}

\DeclareMathOperator*{\subjto}{\,subject\,to\,}

\definecolor{light-gray}{gray}{0.75}

\newtheorem{mydef}{Definition}
\newtheorem*{require}{Requirements}
\newcommand{\Matlab}{$\text{Matlab}^{\circledR}~$}
\newcommand{\IntelProc}{$\text{Intel}^{\circledR}~\text{Core}\texttrademark~\text{i7}~$}
\newcommand{\ie}{\emph{i.e.~}}

\begin{document}
%
\title{A modeling and algorithmic framework for (non)social (co)sparse audio restoration}
%
%
%

\author{C. Gaultier\thanks{C. Gaultier, N. Bertin and R. Gribonval are with Univ Rennes, Inria, CNRS, IRISA. While preparing this work S. Kiti\'c was with Technicolor R\&D.}, 
        N. Bertin, 
        S. Kiti\'c, 
        R. Gribonval, 
} 

\maketitle

\begin{abstract}
We propose a unified modeling and algorithmic framework for audio restoration problem. 
It encompasses analysis sparse priors as well as more classical synthesis sparse priors, and regular sparsity as well as various forms of structured sparsity embodied by shrinkage operators (such as social shrinkage). The versatility of the framework is illustrated on two restoration scenarios: denoising, and declipping. Extensive experimental results on these scenarios highlight both the speedups of 20\% or even more offered by the analysis sparse prior, and the substantial declipping quality that is achievable with both the social and the plain flavor. While both flavors overall exhibit similar performance, their detailed comparison displays distinct trends depending whether declipping or denoising is considered.
\end{abstract}


%
\IEEEpeerreviewmaketitle


\section{Introduction and Motivations}\label{sec:Intro}

\IEEEPARstart{W}{ether} originating from the acquisition, the analog-to-digital conversion or any other later processing step, distortion in audio signals induces unwanted effects and has a negative impact on application performance. Distortion may consist in clipping, packet loss or additive noise. The resulting degraded Signal-to-Noise ratio (SNR) directly affects speech understanding, listening comfort and automated tasks like speech recognition or signal classification.
Audio enhancement aims at restoring such distorted signals. Typical approaches to signal restoration or reconstruction rely on  spectral subtraction~\cite{boll1979spectral}, autoregressive or statistical models~\cite{mcaulay1980speech}, and lately, neural networks~\cite{xu2014speechdnn}. Recently, a body of work popularized the explicit formulation of the enhancement task as an \textit{inverse problem}, and accordingly, the idea of solving it through the use of a time-frequency \textit{sparse regularization} \cite{Mallat99Wavelet,yu2008audio,kiti:tel-01237323}.

\subsection{Analysis {\em vs} Synthesis}

The {\em sparse synthesis model} assumes that the signal of interest $\vect{x}$ is built from a linear combination of atoms aggregated in a large dictionary $\mtrx{D}$. We could more precisely write

\begin{equation}
	\label{eq:SparseSynthesis}
	\vect{x} = \mtrx{D}\vect{z}
\end{equation}  

\noindent with $\vect{x}\in \Rn{L}$ the time domain signal, $\mtrx{D}\Cset{L}{S}$ the dictionary and $\vect{z} \in \Cn{S}$ a sparse representation of the vector $\vect{x}$.

While such synthesis approaches comprise a vast majority of the sparsity-based time-frequency regularization techniques, it has been demonstrated recently \cite{kiti:tel-01237323,kitic2015sparsity} that the {\em analysis sparse model}, which assumed name is cosparse model, can reveal more advantageous, in particular in terms of computational cost. 
Instead of estimating a sparse representation $\vect{z}$ of the signal $\vect{x}$ through the sparse synthesis model, 
the rationale of the cosparse model is to estimate the signal $\vect{x}$ itself assuming that
\begin{equation}
	\label{eq:SparseAnalysis}
	\vect{z} = \mtrx{A}\vect{x}
\end{equation}

\noindent is sparse with $\mtrx{A} \Cset{P}{L}$ called the analysis operator. The two models are equivalent when $\ing{P}=\ing{S}=\ing{L}$ and $\mtrx{A}\mtrx{D} = \mtrx{I}$.

\begin{figure}[!htbp]
	\centering
	\subfloat[Tonal music]{\includegraphics[width=4.3cm]{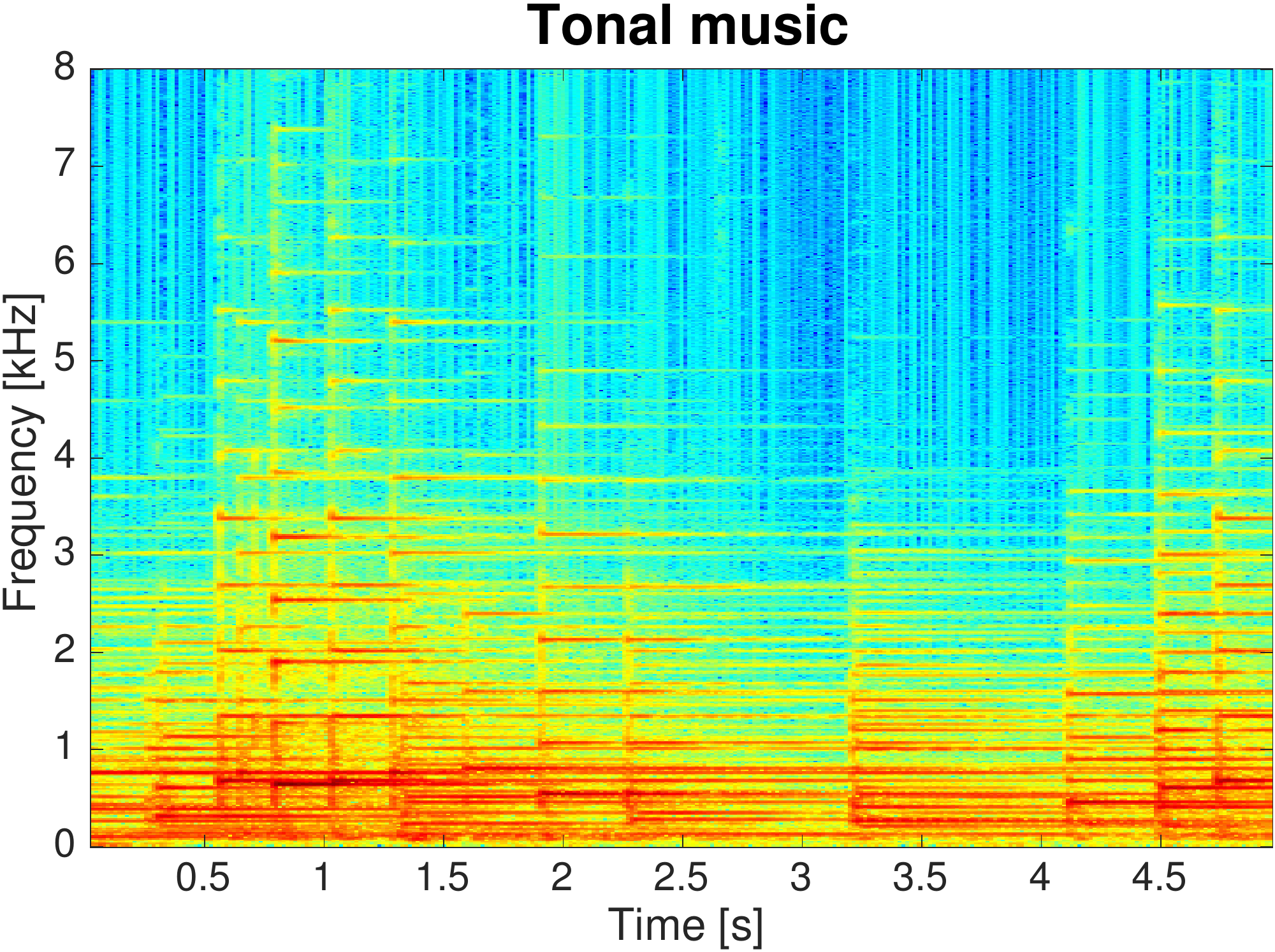}%
		\label{fig:TonalSpec}}
	\hfil
	\subfloat[Transient music]{\includegraphics[width=4.3cm]{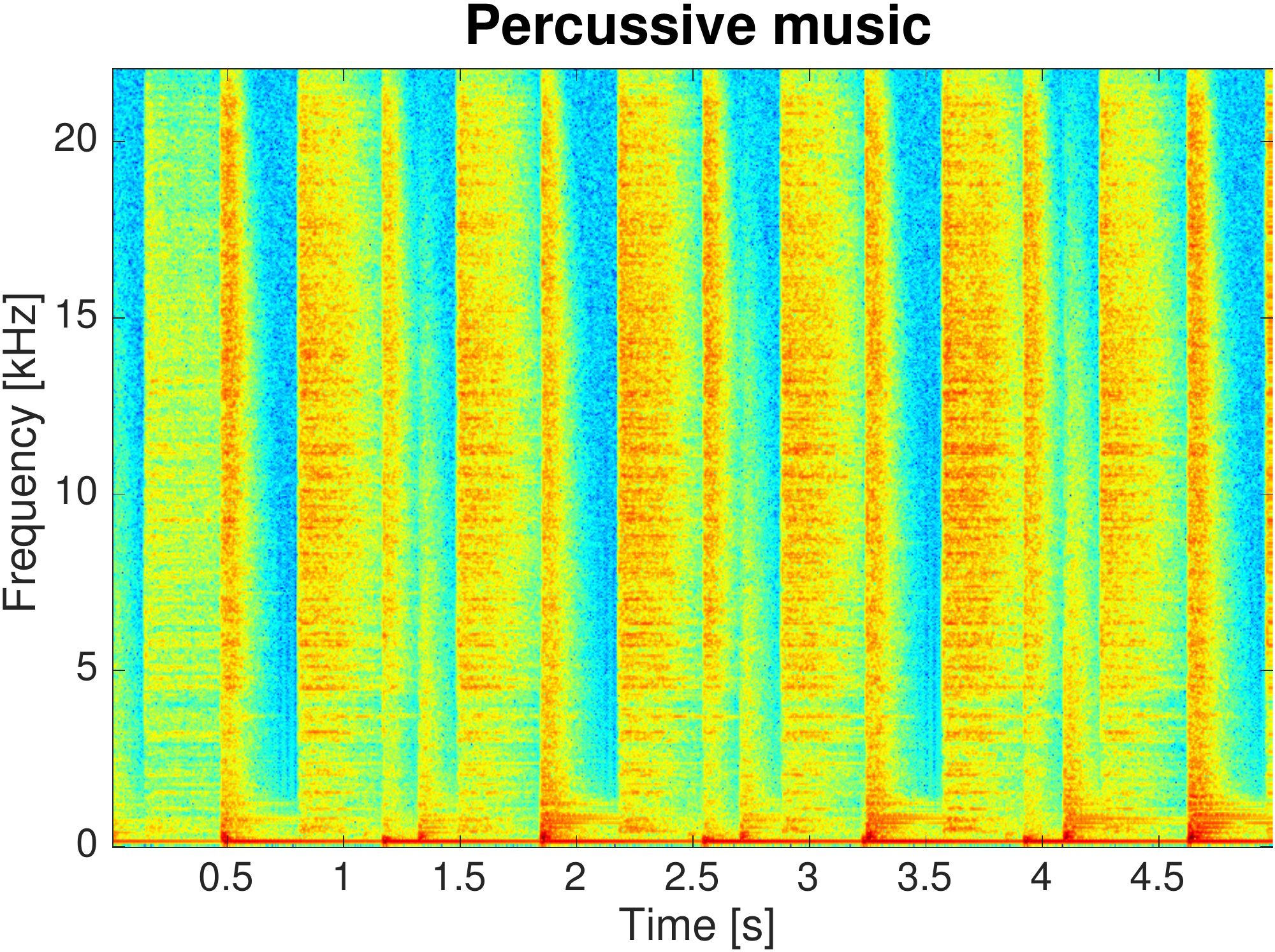}%
		\label{fig:TransientSpec}}
	\hfil
	\caption{Musical excerpts spectrograms}
	\label{fig:CleanSpectrograms}
\end{figure}

\subsection{Structured Sparsity}

Structured forms of sparsity such as group sparsity \cite{jenatton2011structured} or social sparsity \cite{kowalski2013social} have emerged as useful refinements of the above mentioned techniques to take into account the typical time-frequency patterns of audio signals, as illustrated on Figures \ref{fig:TonalSpec} and \ref{fig:TransientSpec}. 
Figure \ref{fig:TonalSpec} displays the spectrogram of a tonal musical excerpt, where high energy coefficients are structured across time reflecting the strong presence of harmonics. Figure \ref{fig:TransientSpec} displays the spectrogram of a percussive music sample, where the dominant coefficients gather across frequency due to transients and beats.

\subsection{Contributions}
In this paper, we introduce a new joint modeling and algorithmic framework encompassing sparse and cosparse models as well as plain or structured sparsity for time-frequency audio restoration. 
The versatility of the framework is illustrated on two audio reconstruction tasks: declipping and denoising\footnote{Audio examples are available at \url{https://project.inria.fr/spare/}}. This is achieved through a unique adaptive algorithmic structure designed to encompass independently the modeling variations or the specific features of different audio enhancement tasks.

The paper is organized as follows. Section \ref{sec:Framework} introduces the general algorithmic framework. Section~\ref{sec:Denoising} (resp. Section~\ref{sec:Declipping}) instantiates this framework for denoising (resp. declipping) including extensive experimental results. On both scenarios, while the analysis and synthesis flavors yield almost identical restoration performance, significant speedups are obtained with the analysis one, hence its performance is studied in more details. For declipping, it yields significant quality improvements for speech and most music styles. The performance of plain {\em vs} social cosparsity is often comparable but significantly distinct trends are observed between the denoising and declipping scenarii when varying the input degradation level. 
The final section collects last notes and suggests future insights.

\subsection{Notations}

In the following, lower-case Greek symbols ($\varepsilon$) stands for scalar constant. Lower-case sans serif font ($\ing{i}$) denotes an integer. Lower-case bold font ($\vect{v}$) expresses a vector and upper-case ($\mtrx{V}$) a matrix. $\vect{v}_{\ing{i}}$ is an $\ing{i}$\textsuperscript{th} element of a vector and $\iter{\vect{v}}{i}$ an $\ing{i}$\textsuperscript{th} iterate. $\Theta$ is used for a set. $\mathcal{O}$ stands for a non-linear operator and $F$ a functional. $\mtrx{V}_{(\ing{i}\ing{j})}$ represents the component of the matrix $\mtrx{V}$ indexed the $\ing{i}$\textsuperscript{th} row and $\ing{j}$\textsuperscript{th} column. Finally, $\htransp{\mtrx{v}}$ denotes the Hermitian transpose of a matrix $\mtrx{v}$. Curved relation symbols ($\preccurlyeq,\succcurlyeq,\prec,\succ$) are used for entry-wise comparisons between matrices. Any other notation will be disambiguated in the text.

\section{General Framework}\label{sec:Framework}

In this section, we present a general framework using either simple sparse modeling (analysis or synthesis based) or structured sparse priors to address reconstruction problems in audio.

\subsection{Modeling Framework}\label{subsec:Modeling}

As for many other problems in audio, we cast the problem of recovering a clean signal $\vect{x}$ from noisy indirect measurements $\vect{y}$ as a linear inverse problem, stated in a frame-based manner.

We consider the matrix $\mtrx{y}_{\ing{n}}$ containing one or more windowed frames of $\ing{L}$ samples from the observed signal $\vect{y}$. The problem of audio reconstruction is to estimate the original clean signal frames, similarly gathered in a matrix $\mtrx{x}_\ing{n}$. 

This can be expressed as an inverse problem, which is usually ill-posed, hence the need to regularize it thanks to prior knowledge on the signal to recover. Such knowledge can be to assume that the signal to recover admits a time-frequency representation endowed with some form of sparsity. 

\subsubsection{Analysis and synthesis sparse modeling}

We consider one frame at a time, i.e. $\mtrx{x}_{\ing{n}} \Rset{L}{1}$, and $\mtrx{z}_{\ing{n}}$ an approximate frequency representation of $\mtrx{x}_{\ing{n}}$ assumed to have few significant coefficients. The assumed relation between $\mtrx{x}_{\ing{n}}$ and $\mtrx{z}_{\ing{n}}$ depends on the type of sparse model:\\
~\\
	\begin{tabular}{m{0.45\columnwidth}|m{0.45\columnwidth}}
	
		\multicolumn{1}{c|}{\textbf{Analysis sparse model}}&\multicolumn{1}{c}{\textbf{Synthesis sparse model}}                                                            \\ 
		$\mtrx{A} \Cset{P}{L}, \ing{P} \geq \ing{L}$ & $\mtrx{D} \Cset{L}{S}, \ing{S} \geq\ing{L}$                                                                                                                                                            \\
		$\mtrx{z}_{\ing{n}} \simeq \mtrx{A}\mtrx{x}_{\ing{n}}, \mtrx{z}_{\ing{n}} \Cset{P}{1}$&$\mtrx{D}\mtrx{z}_{\ing{n}} \simeq \mtrx{x}_{\ing{n}}, \mtrx{z}_{\ing{n}} \Cset{S}{1}$
		\\
		$\norm{\mtrx{z}_{\ing{n}}}{0}\ll\ing{P}$;&$\norm{\mtrx{z}_{\ing{n}}}{0}\ll\ing{S}$.
	\end{tabular}
~\\\\
The matrix $\mtrx{A}$ (resp. $\mtrx{D}$) embodies a forward (resp. backward) frequency transform (e.g. DCT or DFT), possibly made redundant with zero-padding.
\subsubsection{From plain to structured sparse modeling}
The above-described models -- which will be denoted as ``plain'' sparse models -- treat separately each frame of the signal. In contrast, ``structured'' sparse modeling introduces dependencies between frequency representations of adjacent time frames.
For this, we consider the matrix $\mtrx{x}_{\ing{n}} \Rset{L}{(2b+1)}$ which columns are the frames of the original signal indexed by $[\ing{n}-\ing{b}, \ing{n}+\ing{b}]$, and $\mtrx{Z}_{\ing{n}}$ a matrix which columns are frequency representations of these frames. In other words, this matrix is a \emph{time-frequency} representation of the underlying audio signal. 

In structured sparse models, the assumed relation between $\mtrx{Z}_{\ing{n}}$ and $\mtrx{X}_{\ing{n}}$ becomes:\\
~\\
	\begin{tabular}{m{0.48\columnwidth}|m{0.48\columnwidth}}
		\multicolumn{1}{c|}{\textbf{Structured analysis model}}&\multicolumn{1}{c}{\textbf{Structured synthesis model}}                                                            \\ 
		$\mtrx{A} \Cset{P}{L}, \ing{P} \geq \ing{L}$ & $\mtrx{D} \Cset{L}{S}, \ing{S} \geq\ing{L}$                                                                                                                                                            \\
		$\mtrx{Z}_{\ing{n}} \simeq \mtrx{A}\mtrx{X}_{\ing{n}}, \mtrx{Z}_{\ing{n}} \Cset{P}{(2b+1)}$&$\mtrx{D}\mtrx{Z}_{\ing{n}} \simeq \mtrx{X}_{\ing{n}}, \mtrx{Z}_{\ing{n}} \Cset{S}{(2b+1)}$
		\\
		$\norm{\mtrx{z}_{\ing{n}}}{0}\ll\ing{P}\times\ing{(2b+1)}$&$\norm{\mtrx{z}_{\ing{n}}}{0}\ll\ing{S}\times\ing{(2b+1)}$
		\\
		$\mtrx{z}_{\ing{n}}$ is ``structured''; &$\mtrx{z}_{\ing{n}}$ is ``structured''.
	\end{tabular}
~\\

In this paper, to instantiate the notion of ``structured'' sparse modeling, we rely on the concept of \emph{social sparsity} \cite{kowalski2013social}, \ie we assume that the indices of non-zero coefficients in $\mtrx{Z}_{\ing{n}}$ are organized according to some known time-frequency \emph{patterns} that will be illustrated in Section~\ref{subsec:SparseOperators}. 

\underline{\emph{Remark:}}~for the special case where $\ing{b}=0$, both $\mtrx{X}_{\ing{n}}$ and $\mtrx{Z}_{\ing{n}}$ have a single column, and the social sparse modeling collapses to the plain sparse assumption.

\subsection{Algorithmic Framework}\label{subsec:Algorithm}

Given a distorted matrix of observations $\mtrx{Y}_{\ing{n}}$, our goal is to find means to recover an estimate $\hat{\mtrx{x}}_{\ing{n}}$ of the frames $\mtrx{x}_{\ing{n}}$ of the original signal. For this, one seeks $\hat{\mtrx{x}}_{\ing{n}}$ that satisfies:
\begin{itemize}
\item a data fidelity constraint with respect to $\mtrx{Y}_{\ing{n}}$, according to some distortion model (additive noise, clipping...);
\item the modeling constraints described above.
\end{itemize}
This is the spirit of the algorithmic framework we develop. It relies on two components: 
\begin{itemize}
\item a \emph{generalized projection} onto the data-fidelity constraint;
\item a \emph{shrinkage} enforcing (structured) sparsity.
\end{itemize}

These are combined into an iterative algorithm analog to Douglas-Rachford (DR) splitting \cite{eckstein1992douglas}. 
This generic algorithmic framework can be instantiated in different ways depending on the target application. Two specific examples will be thoroughly described in Sections~\ref{sec:Denoising}  (denoising) and Section~\ref{sec:Declipping} (declipping). 

  The first component is the generalized projection:
 \begin{mydef}[Generalized projection]
Let $\Theta$ be a nonempty \emph{convex} set, and $\mtrx{M}$ be a full column rank matrix. Given a time-frequency matrix $\mtrx{Z}$, we denote $\mathcal{P}_{\Theta,\mtrx{M}}(\mtrx{Z})$ the (unique) solution of the following optimization problem:
 	\begin{equation}\label{eq:ConstrainedProjection}
 	\minim_{\mtrx{W} \in \Theta} \norm{\mtrx{M} \mtrx{W} - \mtrx{Z}}{\text{F}}. 
 	\end{equation} 
 \end{mydef}
 The computation of this projection for some particular choices of constraint set $\Theta$ and matrix $\mtrx{M}$ will be discussed in due time.
 
   The second component is the shrinkage operator.
Intuitively, this operator gives an output which is ``decreased'' in a certain sense, with respect to its input argument, hence somewhat promoting sparsity. Although we will not formally exploit it for any convergence analysis, we also recall below the notion of shrinkage, also called ``thresholding rule'' \cite{kowalski2014thresholding}.
 \begin{mydef}[Shrinkage] $\mathcal{S}(\cdot)$, is a shrinkage if:
 	\begin{enumerate}
 		\item $\mathcal{S}(\cdot)$ is an odd function;
		\item $0 \leq \mathcal{S}(x) \leq x$, for all $x \in \mathbb{R}^+$.
 		\item $(\mathcal{S}(\cdot))_+$ is nondecreasing on $\mathbb{R}^+$ and $\lim_{x \rightarrow +\infty} (\mathcal{S}(x))_+ = +\infty$, where	$(\cdot)_+:= \max(\cdot,0)$.
 	\end{enumerate}
 \end{mydef}
When applied to a (time-frequency) matrix, and written $\mathcal{S}(\mtrx{Z})$, shrinkage is applied entry-wise. 

In a concrete setting, the following is required to instantiate the framework:
\begin{require}~\\
\begin{itemize}
\item a convex set $\Theta$ and a matrix $\mtrx{M}$ embodying the data fidelity constraint and the domain (time or frequency) in which it is specified;
\item a parameterized family of shrinkages $\{\mathcal{S}_{\mu}(\cdot)\}_{\mu}$, where the amount of shrinkage is controlled by $\mu$: in the extreme cases $\mathcal{S}_0(\mtrx{Z}) = \mtrx{Z}$ and $\mathcal{S}_{\infty}(\mtrx{Z}) = \mtrx{0}$; 
\item a rule $F: \mu \mapsto F(\mu)$ to update the amount of shrinkage across iterations, and an initial $\iter{\mu}{\ing{0}}$;
\item an initial estimate $\iter{\mtrx{Z}}{0}$ of the seeked time-frequency representation;
\item stopping parameters $\beta$ and $\ing{i}_{\max}$.
\end{itemize}
\end{require}
The proposed generic algorithm is described in Algorithm~\ref{alg:AbstractAlgo}.
 \begin{algorithm}
 	
 	\caption{Generic Algorithm: $\mathcal{G}$}
 	\label{alg:AbstractAlgo}
 	
 	\begin{algorithmic} 
\REQUIRE $\Theta, \mtrx{M}, \{\mathcal{S}_{\mu}(\cdot)\}_{\mu}, \iter{\mu}{\ing{0}}, F(\cdot), \iter{\mtrx{Z}}{0}, \beta, \ing{i}_{\max}$
\STATE $\iter{\mtrx{U}}{0}=\mtrx{0}$;
\FOR{$\ing{i} = 1 \text{ to } \ing{i}_{\max}$}
\STATE $\iter{\mtrx{w}}{i} =  \mathcal{P}_{\Theta,\mtrx{M}}(\iter{\mtrx{z}}{i-1} - \iter{\mtrx{u}}{i-1})$
\STATE $\iter{\mtrx{z}}{i} =  \mathcal{S}_{\iter{\mu}{i-1}} \left( \mtrx{M} \iter{\mtrx{w}}{i} + \iter{\mtrx{u}}{i-1} \right)$
\IF{$\frac{\norm{\mtrx{M}\iter{\mtrx{w}}{i} - \iter{\mtrx{Z}}{i}}{\text{F}} }{\norm{\mtrx{M}\iter{\mtrx{w}}{i}}{\text{F}}} \leq \beta$}
\STATE $\text{terminate}$
\ELSE
\STATE $\iter{\mtrx{u}}{i} = \iter{\mtrx{u}}{i-1} +  \mtrx{M} \iter{\mtrx{w}}{i} - \iter{\mtrx{z}}{i}.$
\STATE $\iter{\mu}{i} = F(\iter{\mu}{i-1})$
\ENDIF
\ENDFOR
\RETURN $\iter{\mtrx{w}}{\ing{i}}$\ [and optionally $\iter{\mu}{i}$, $\iter{\mtrx{z}}{i}$] 
	\end{algorithmic}
\end{algorithm}

The notation $\iter{\mtrx{Z}}{i}$ highlights that the corresponding variable is in any use-case a sparse/structured time-frequency representation. The variable $\iter{\mtrx{u}}{i}$ is an intermediate time-frequency ``residual'' variable typical of ADMM / Douglas-Rachford. At iteration $\ing{i}$, an estimate of $\mtrx{Z}_{\ing{n}}$ is $\iter{\hat{\mtrx{Z}}}{i} := \iter{\mtrx{z}}{i-1} - \iter{\mtrx{u}}{i-1}$. The interpretation of the other variables is use-case dependent:
\begin{itemize}
\item {\bf analysis flavor:} $\mtrx{M} := \mtrx{A}$ is the frequency analysis operator; $\iter{\mtrx{w}}{i}$ is an estimate of the time frames $\mtrx{X}_{\ing{n}}$, that satisfies the time-domain data-fidelity constraint $\Theta$ while being closest to $\iter{\hat{\mtrx{Z}}}{i}$  \emph{in the time-frequency domain}; the algorithm outputs a time-domain estimate.
\item {\bf synthesis flavor:}  $\mtrx{M} := \mtrx{I}$; $\iter{\mtrx{w}}{i}$ is a time-frequency estimate of $\mtrx{Z}_{\ing{n}}$; the data-fidelity constraint $\Theta$ is \emph{expressed in the time-frequency domain}; the algorithm outputs a time-frequency estimate, from which it is possible to get a time-domain estimate by synthesis $\hat{\mtrx{x}}_{\ing{n}} := \mtrx{D}\iter{\mtrx{W}}{i}$.
\end{itemize}
Due to the expression of $\Theta$ respectively in the time domain and the time-frequency domain, the analysis and synthesis flavors will have different computational properties (for declipping mainly) as will be further discussed in Sections~\ref{sec:expe_denoising} and~\ref{sec:expe_declipping}.
Algorithm \ref{alg:AbstractAlgo} is summarized as a generalized procedure $\mathcal{G}(
\Theta, \mtrx{M}, \{\mathcal{S}_{\mu}\}_{\mu}, \iter{\mu}{\ing{0}}, F, \iter{\mtrx{Z}}{0}, \beta, \ing{i}_{\max})$.

\subsection{Shrinkages for (social) sparsity}
\label{subsec:SparseOperators}

As noted in the requirements, we need to choose the family $\{\mathcal{S}_{\mu}(\cdot)\}_{\mu}$ of shrinkage operators. 

For plain sparsity, either analysis or synthesis, we use the hard-thresholding operator $\mathcal{H}_{\ing{k}} (\mtrx{Z})$ that sets all but the $\ing{k}$ coefficients of largest magnitude in $\mtrx{Z}$ to zero (see e.g. \cite{blumensath2009iterative}).
In the case of analysis sparse modeling with $\mtrx{A} \Cset{P}{L}$ a forward frequency analysis operator
(resp. $\mtrx{D} \Cset{L}{S}$ a dictionary) we set $\mathcal{S}_{\mu} := \mathcal{H}_{\ing{P} - \mu}$ (resp. $\mathcal{S}_{\mu} := \mathcal{H}_{\ing{S} - \mu}$), for $\mu \in \mathbb{N}^+$, $0 \leq \mu \leq P$ (resp. $0 \leq \mu \leq S$).

For social sparsity (again, either analysis or synthesis), we choose the Persistent Empirical Wiener (PEW) operator \cite{kowalski2014thresholding} successfully used in \cite{siedenburg2014audio} for audio declipping. 
This shrinkage promotes specific local time-frequency structures around each time-frequency point. Its specification explicitly requires choosing a time-frequency pattern described as a matrix $\Gamma \Rset{(\ing{2F+1})}{(\ing{2T+1})}$ with binary entries. 

\begin{figure}[!htbp]
	\centering
	\includegraphics[width=0.66\columnwidth
	]{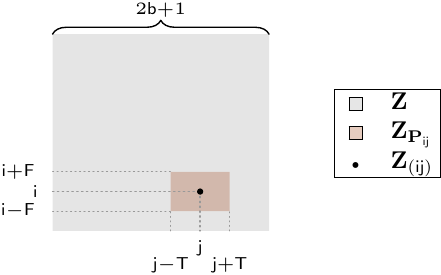}
	\caption{Schematic representation of patch extraction from matrix $\mtrx{Z}$\label{fig:Z}}
\end{figure}

Rows of $\Gamma$ account for the frequency dimension and columns for the time dimension, in \emph{local time-frequency coordinates}. 
Let $\mtrx{Z}_{\ing{n}} \Cset{\ing{L}}{\ing{(2b+1)}}$ be a time-frequency representation. For clarity of presentation, we will now omit the $\ing{n}$ index and simply denote $\mtrx{Z}$ (and similarly for $\mtrx{X}$, $\mtrx{Y}$ later on when needed). 
As illustrated on Figure~\ref{fig:Z}, consider $\ing{ij}$ the coordinates of a time-frequency point in $\mtrx{Z}$ and $\mtrx{P}_{\ing{ij}} := [\ing{i}-F,\ing{i}+F] \times [\ing{j}-T,\ing{j}+T]$ the indices corresponding to a time-frequency patch of size $(\ing{2F+1}) \times (\ing{2T+1})$ centered in $\ing{ij}$. The matrix $\mtrx{Z}_{\mtrx{P}_{\ing{ij}}} \Cset{(\ing{2F+1})}{(\ing{2T+1})}$ is extracted from $\mtrx{Z}$ on these indices, with mirror-padding on the borders if needed. 

Now that we have expressed how $\mtrx{Z}, \mtrx{Z}_{\mtrx{P}_{\ing{ij}}}$ and indexes are organized, we can define PEW using $\circ$ to denote the Hadamard product:

\begin{equation}\label{eqPEW}
\mathcal{S}^{\text{PEW}}_{\mu}(\mtrx{Z} | \Gamma)_{(\ing{i}\ing{j})} := \mtrx{Z}_{(\ing{i}\ing{j})} \cdot \left(1 - \frac{\mu^2}{\norm{ \mtrx{Z}_{\mtrx{P}_{\ing{ij}}} \circ \Gamma}{2}^2 } \right)_+.
\end{equation}
Since $\norm{\mtrx{Z}_{\mtrx{P}_{\ing{ij}}} \circ \Gamma}{2}^2$ is the energy of $\mtrx{Z}$ restricted to a time-frequency neighborhood of $\ing{ij}$ of shape specified by $\Gamma$, the left hand side is zero as soon as this energy falls below $\mu^{2}$. As such, PEW shrinkage effectively promotes structured sparsity.

\begin{figure}[!htbp]
	\centering
	\subfloat[$\Gamma_{\ing{1}}$]{\includegraphics[height=7mm]{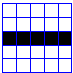}}%
	\hfil
	\subfloat[$\Gamma_{\ing{2}}$]{\includegraphics[height=7mm]{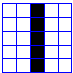}}%
	\hfil
	\subfloat[$\Gamma_{\ing{3}}$]{\includegraphics[height=7mm]{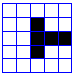}}%
	\hfil
	\subfloat[$\Gamma_{\ing{4}}$]{\includegraphics[height=7mm]{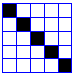}}%
	\hfil
	\subfloat[$\Gamma_{\ing{5}}$]{\includegraphics[height=7mm]{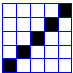}}%
	\hfil
	\subfloat[$\Gamma_{\ing{6}}$]{\includegraphics[height=7mm]{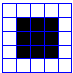}}%
	\caption{Extended set of time-frequency neighborhoods used for music\label{fig:Stencils}}
\end{figure}

\begin{figure}[!htbp]
	\centering
	\subfloat[$\Gamma_{\ing{1}}$]{\includegraphics[height=7mm]{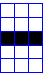}}%
	\hfil
	\subfloat[$\Gamma_{\ing{2}}$]{\includegraphics[height=7mm]{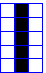}}%
	\hfil
	\subfloat[$\Gamma_{\ing{3}}$]{\includegraphics[height=7mm]{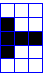}}%
	\hfil
	\subfloat[$\Gamma_{\ing{4}}$]{\includegraphics[height=7mm]{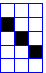}}%
	\hfil
	\subfloat[$\Gamma_{\ing{5}}$]{\includegraphics[height=7mm]{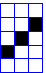}}%
	\hfil
	\subfloat[$\Gamma_{\ing{6}}$]{\includegraphics[height=7mm]{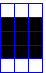}}%
	\caption{Extended set of time-frequency neighborhoods used for music\label{fig:SpeechStencils}}
\end{figure}

Examples of patterns for music  are given in Figure~\ref{fig:Stencils} and for speech in Figure~\ref{fig:SpeechStencils}. They are similar but at different time scales, given the different scales of stationarity in speech and music. The structures embedded in these patterns have various properties: $\Gamma_{\ing{1}}$, with a frequency localized and time-spread support, will emphasize tonal content; vice-versa, $\Gamma_{\ing{2}}$ will emphasize transients and attacks; $\Gamma_{\ing{3}}$ is designed \cite{siedenburg2012audio} to avoid pre-echo artifacts; patterns $\Gamma_{\ing{4}}$ and $\Gamma_{\ing{5}}$ are introduced to stress tonal transitions; finally, $\Gamma_{\ing{6}}$ serves as a default pattern 
when no particular structure is identified.\\

\underline{\emph{Remarks:}} ~Here the subscript index $\ing{k}$ for each time-frequency pattern $\Gamma_{\ing{k}}, \ing{k}\in\{1..6\}$ is not a time frame index but counts the patterns within the collection.\\
On Fig.~\ref{fig:Stencils} the total time span for each $\Gamma$ is 320 ms. On Fig.~\ref{fig:SpeechStencils} the total time span reduces to 96 ms.

\captionsetup[subfigure]{labelformat=empty}

\captionsetup[subfigure]{labelformat=parens}

Now that we described all the useful modeling and algorithmic tools used in our audio reconstruction scheme, sections~\ref{sec:Denoising} and~\ref{sec:Declipping} will instantiate the enhancement procedure respectively for denoising and declipping and present experimental results.

\section{Audio Denoising Use Case\label{sec:Denoising}}

Denoising is one of the most intensively studied inverse problems in audio signal processing. Whether it originates from the environment or the microphones, noise is an inevitable (and, usually, undesirable) component of audio recordings, calling for a denoising block in signal processing pipelines for applications such as speech recognition, sound classification, and many others.

A specific degraded model in case of additive noise writes:
\begin{equation}
	\label{eq:NoiseModel}
\mtrx{y}_\ing{n} = \mtrx{x}_\ing{n} + \mtrx{e}_\ing{n},
\end{equation}
with $\mtrx{e}_{\ing{n}}$ often modeled as white Gaussian noise of fixed variance $\sigma^2$ in the frame(s) around frame $\ing{n}$. 

\subsection{Generalized projections for the denoising problem}

A natural expression of the data-fidelity constraint is of the form $\norm{\hat{\mtrx{X}}-\mtrx{Y}}{\text{F}} \leq \varepsilon$ for some $\varepsilon$. Heuristics to choose $\varepsilon$ given an estimated variance $\sigma^{2}$ will be discussed in Section~\ref{sec:expe_denoising}.

In the analysis setting, with $\mtrx{M} := \mtrx{A}$, the data-fidelity constraint yields $\Theta := \{ \mtrx{W} \ |\ \norm{\mtrx{W}-\mtrx{Y}}{\text{F}} \leq \varepsilon\}$. In the synthesis setting, with $\mtrx{M} := \mtrx{I}$, we set $\Theta := \{ \mtrx{W} \ |\ \norm{\mtrx{D}\mtrx{W}-\mtrx{Y}}{\text{F}} \leq \varepsilon\}$. These choices hold both for plain and social versions.

In the analysis setting, assuming $\htransp{\mtrx{A}}\mtrx{A} = \mtrx{I}$, the desired projection can be expressed in closed-form as:
\begin{equation}\label{eq:SocCosparseDenoiseClosedForm}
	\mathcal{P}_{\Theta,\mtrx{M}}(\mtrx{Z}) = \htransp{\mtrx{A}}\mtrx{Z} - \left(\tfrac{\norm{\htransp{\mtrx{A}}\mtrx{Z} -\mtrx{y}}{\text{F}}-\varepsilon}{\norm{\htransp{\mtrx{A}}\mtrx{Z} -\mtrx{y}}{\text{F}}}\right)_{+}\cdot\left(\htransp{\mtrx{A}}\mtrx{Z} -\mtrx{y}\right)
\end{equation}
as shown in Appendix~\ref{app:ProxDenoising} for the more general case $\htransp{\mtrx{A}}\mtrx{A} \propto \mtrx{I}$.

Hence, in this case, the cost of computing  the generalized projection is dominated by matrix-vector products with $\htransp{\mtrx{M}}$. When this can be done with a fast transform, the analysis flavor has low complexity. When complex transforms are used, to ensure the estimate is real, we replace $\htransp{\mtrx{m}}\mtrx{z}$ in~\eqref{eq:SocCosparseDenoiseClosedForm} with $\Re(\htransp{\mtrx{m}}\mtrx{z})-\Im(\htransp{\mtrx{m}}\mtrx{z})$ where $\Re(\cdot)$ and $\Im(\cdot)$ respectively denote the real and imaginary part.

For the synthesis version, assuming $\mtrx{D}\htransp{\mtrx{D}} = \mtrx{I}$, the generalized projection again reduces algebraically to the closed-form expression:
\begin{align}\label{eq:SocSparseDenoiseClosedForm}
\mathcal{P}_{\Theta,\mtrx{M}}(\mtrx{Z})
&	= \mtrx{Z} - \left(\tfrac{\norm{\mtrx{D}\mtrx{Z} -\mtrx{y}}{\text{F}}-\varepsilon}{\norm{\mtrx{D}\mtrx{Z} -\mtrx{y}}{\text{F}}}\right)_{+} \cdot\htransp{\mtrx{D}}(\mtrx{D}\mtrx{Z} - \mtrx{y}).
\end{align}

\subsection{Algorithms for the denoising inverse problem}\label{subsec:SolveDenoise}
We are now ready to instantiate the general algorithm $\mathcal{G}$  in the different cases.
\subsubsection{Plain sparse audio denoisers}

For both the analysis and the synthesis version, we instantiate the general algorithm $\mathcal{G}$ with the choices summarized in Table~\ref{tab:SparseDenoise}.
\begin{table}[h!]
	\caption{Parameters of Algorithm~\ref{alg:AbstractAlgo} for the Plain Sparse Denoiser \label{tab:SparseDenoise}}
	\centering
		\begin{tabular}{l|l}
			\multicolumn{1}{c|}{\textbf{Analysis}}                                                                                                                                 & \multicolumn{1}{c}{\textbf{Synthesis}}                                                                                                                                                                   \\ 
			\\
			$\Theta = \left\{ \mtrx{w} \mid \norm{\mtrx{w} - \mtrx{y}}{\text{2}} \leq \varepsilon \right\}$ & 
			$\Theta = \left\{ \mtrx{w} \mid \norm{\mtrx{D} \mtrx{w} - \mtrx{y}}{\text{2}} \leq \varepsilon \right\}$
			\\
			$\mtrx{M} = \mtrx{A} \Cset{P}{L}, \ing{P} \geq \ing{L}$                                                                                                        & $\mtrx{M} = \mtrx{I} \Cset{L}{L}$,                                                                                                                                                            \\
			$\mathcal{S}_{\mu} (\cdot)= \mathcal{H}_{\ing{P} - \mu} (\cdot)$                                                                              & $\mathcal{S}_{\mu} (\cdot)= \mathcal{H}_{\ing{S} - \mu} (\cdot)$,                                                                                                                                        \\
			$\iter{\mu}{0} = \ing{P}-1$
			&
			$\iter{\mu}{0} = \ing{S}-1$
			\\
			$F : \mu \mapsto \mu - 1$
			&
			$F : \mu \mapsto \mu - 1$
			\\
			$\iter{\mtrx{z}}{0} = \mtrx{A}\mtrx{y}$
			&
			$\iter{\mtrx{z}}{0} = \htransp{\mtrx{D}}\mtrx{y}$
		\end{tabular}%
~\\
\end{table}

The choice of function $F$ and initialization $\iter{\mu}{0}$ means that we start with a small number $\ing{P}-\iter{\mu}{0} = 1$ (resp. $\ing{S}-\iter{\mu}{0} = 1$) of nonzero coefficients for the sparse constraint which we relax gradually as iterations progress. 

The practical choice of the stopping parameter $\beta$ is driven by a compromise between quality and computation time and we will specify the values used in the experimental section.

Algorithm \ref{alg:AbstractAlgo} with these parameters yields:
\[
	\hat{\mtrx{w}} := 
	\mathcal{G}(
\Theta, \mtrx{M}, \{\mathcal{S}_{\mu}(\cdot)\}_{\mu}, \iter{\mu}{\ing{0}}, F, \iter{\mtrx{Z}}{0}, \beta, \ing{i}_{\max}
).
\]
 
For the analysis version $\hat{\mtrx{x}} := \hat{\mtrx{w}}$, while for the synthesis version $\hat{\mtrx{x}} := \mtrx{D}\hat{\mtrx{w}}$.

\subsubsection{Social sparse audio denoisers}\label{sec:SocialSparseDenoisers}

For the social sparse versions of the denoising method, we change the sparsifying operator from $\mathcal{H}_{\ing{P} - \mu} (\cdot)$ to $\NLOp{S}_{\mu}^{\text{PEW}} (\cdot|\Gamma)$, as well as the update rule which becomes $F_{\alpha}: \mu \mapsto \alpha \mu$. The initial value $\iter{\mu}{0}$ may depend on the pattern $\Gamma$ and will be specified in Section~\ref{sec:expe_denoising}.
The resulting parameters are summarized in Table~\ref{tab:SocialSparseDenoise}.
\begin{table}[h!]
	\caption{Parameters of Algorithm~\ref{alg:AbstractAlgo} for the Social Sparse Denoiser \label{tab:SocialSparseDenoise}}
	\centering
		\begin{tabular}{l|l}
			\multicolumn{1}{c|}{\textbf{Analysis}}                                                                                                                                                                                                                                                               & \multicolumn{1}{c}{\textbf{Synthesis}}                                                                                                                                                                   \\ 
			\\
			$\Theta = \left\{ \mtrx{w} \mid \norm{\mtrx{w} - \mtrx{y}}{\text{2}} \leq \varepsilon \right\}$ & 
			$\Theta = \left\{ \mtrx{w} \mid \norm{\mtrx{D} \mtrx{w} - \mtrx{y}}{\text{2}} \leq \varepsilon \right\}$
			\\
			$\mtrx{M} = \mtrx{A} \Cset{P}{L}, \ing{P} \geq \ing{L}$                                                                                                                                                                                                                   & $\mtrx{M} = \mtrx{I} \Cset{L}{L}$,                                                                                                                                                            \\
			$\mathcal{S}_{\mu} (\cdot)= \mathcal{S}^{\text{PEW}}_{\mu}(\cdot|\Gamma)$                                                                                                                                                                & $\mathcal{S}_{\mu} (\cdot)= \mathcal{S}^{\text{PEW}}_{\mu}(\cdot|\Gamma)$,                                                                                                                                        \\
			$\iter{\mu}{0}$: see Section~\ref{sec:expe_denoising}
			&
			$\iter{\mu}{0}$: see Section~\ref{sec:expe_denoising}
			\\
			$F = F_{\alpha}: \mu \mapsto \alpha \mu$
			&
			$F = F_{\alpha}: \mu \mapsto \alpha \mu $
			\\
			$\iter{\mtrx{z}}{0} = \mtrx{A}\mtrx{y}$
			&
			$\iter{\mtrx{z}}{0} = \htransp{\mtrx{D}}\mtrx{y}$
		\end{tabular}%
~\\
\end{table}

A first version of the denoiser works with a \emph{predefined} time-frequency pattern $\Gamma$ and is compactly written as:

\begin{eqnarray*}
	\begin{bmatrix}
		\hat{\mtrx{w}}(\Gamma)\\ 
		\mu(\Gamma)\\ 
		\mtrx{Z}(\Gamma)
	\end{bmatrix} := \mathcal{G}(
	\Theta, \mtrx{M}, \{\mathcal{S}^{\text{PEW}}_{\mu}(\cdot|\Gamma)\}_{\mu}, \iter{\mu}{\ing{0}}, F_{\alpha}, \iter{\mtrx{Z}}{0}, \beta, \ing{i}_{\max}
	).
\end{eqnarray*}

A more adaptive denoiser uses this first version as a building brick to \emph{select} the pattern $\Gamma$ within a prescribed collection. Indeed, in order to get a fully adaptive denoising procedure, we design a method to automatically select the optimal $\Gamma$ for the signal frames at stake. We call this step the ``initialization loop''. It consists in evaluating $\hat{\mtrx{w}}(\Gamma)$ with a small number of iterations (e.g. $\ing{i}_{\max}^{\texttt{small}} = 10$) for different patterns $\Gamma$.

Given a predefined set of time-frequency patterns $\left\{ \Gamma_{\ing{k}}\right\}_{\ing{k}=1}^{\ing{K}}$ and initial threshold values $\iter{\mu}{0}_{\ing{k}}$ that will be specified in Section~\ref{sec:expe_denoising}, one can compute $\hat{\mtrx{w}}_{\ing{k}}:=\hat{\mtrx{w}}(\Gamma_{\ing{k}})$ for $1 \leq \ing{k} \leq \ing{K}$, and similarly $\mu_{\ing{k}} := \mu(\Gamma_{\ing{k}})$ and $\mtrx{Z}_{\ing{k}}:= \mtrx{Z}(\Gamma_{\ing{k}})$. Then, the idea is that the best estimate $\hat{\mtrx{w}}_{\ing{k}}$ should produce a residual with spectrum closest to that of Additive White Gaussian Noise (AWGN), which is by definition flat. Thus, we select the pattern $\Gamma_{\ing{k}^{\star}}$ yielding a residual with time-frequency representation of highest entropy.

For a given $\ing{k}$, we can define the resulting time-frequency residual: $\mtrx{R}_{\ing{k}}  := \mtrx{M}\hat{\mtrx{w}}_{\ing{k}} - \iter{\mtrx{Z}}{0}$. Computing a $\ing{Q}$-bin histogram of the modulus of its entries yields $\hat{\var{p}}$, an empirical probability distribution, which (empirical) entropy is
\begin{equation}
\label{eq:entropy}
\var{e}_{\ing{k}} = - \sum_{\ing{q=1}}^{\ing{Q}} \hat{\var{p}}_{\ing{q}} \log_2  (\hat{\var{p}}_{\ing{q}}).
\end{equation}
A heuristic to choose $\ing{Q}$ is the Herbert-Sturges rule \cite{sturges1926choice}
\begin{equation}
\label{eq:sturges}
\ing{Q} = \lfloor 1 + \log_2 (\# \mtrx{R}_{\ing{k}} ) \rfloor,
\end{equation}
where $\lfloor\cdot\rfloor$ is the floor function and $\# \mtrx{R}_{\ing{k}} = \ing{L} \times (2\ing{b}+1)$ is the number of entries in the matrix $\# \mtrx{R}_{\ing{k}}$. The values considered in the experiments of Section~\ref{sec:expe_denoising} lead to $\ing{Q} \in \{13,15\}$.

Once the best pattern $\Gamma_{\ing{k}^{\star}}$ is chosen as just described, we run Algorithm \ref{alg:AbstractAlgo} with the parameters of Table~\ref{tab:SocialSparseDenoise} and warm-started $\mu^{(0)}$ and $\mtrx{Z}^{(0)}$,  
with a sufficiently large $\ing{i}_{\max}$ (typically $\ing{i}_{\max}^{\texttt{large}}= 10^6$) to get
\[
\hat{\mtrx{w}}:=\mathcal{G}(
\Theta, \mtrx{M}, \{\mathcal{S}^{\text{PEW}}_{\mu}(\cdot|\Gamma_{\ing{k}^{\star}})\}_{\mu}, 
\mu_{\ing{k}^{\star}}, F_{\alpha}, \mtrx{Z}_{\ing{k}^{\star}},  \beta, \ing{i}_{\max}^{\texttt{large}}
).
\] 
The pseudo-code of the adaptive social denoiser for a given block of adjacent frames $\mtrx{Y} \Rset{\ing{L}}{\ing{(2b+1)}}$ is given in Algorithm \ref{alg:AUDASCITY}. Again, for the analysis version $\hat{\mtrx{x}} := \hat{\mtrx{w}}$, while for the synthesis version $\hat{\mtrx{x}} := \mtrx{D}\hat{\mtrx{w}}$.

\begin{algorithm}
	\caption{Adaptive Social Sparse Denoisers 	\label{alg:AUDASCITY}}
	
	\begin{algorithmic} 
		\REQUIRE $\mtrx{Y}$, $\varepsilon$, $\mtrx{A}$ or $\mtrx{D}$, $\left\{\Gamma_{\ing{k}}\right\}_{\ing{k}}$, $\{\iter{\mu_{\ing{k}}}{0}\}_{\ing{k}}$, $\alpha,\beta$, $\ing{i}_{\max}^{\texttt{small}}$, $\ing{i}_{\max}^{\texttt{large}}$
		\STATE set parameters from Table~\ref{tab:SocialSparseDenoise}
		\FORALL{$\ing{k}$}
		\STATE
		\STATE \resizebox{0.45\textwidth}{!}{$\begin{bmatrix}
		\hat{\mtrx{w}}_{\ing{k}}\\ 
		\mu_{\ing{k}}\\ 
		\mtrx{Z}_{\ing{k}}
		\end{bmatrix}:= 
		\mathcal{G}(\Theta, \mtrx{M},\{\mathcal{S}_{\mu}^{\text{PEW}}(\cdot|\Gamma_{\ing{k}})\}_{\mu}, \iter{\mu_{\ing{k}}}{\ing{0}}, F_{\alpha}, \iter{\mtrx{z}}{0}, \beta, \ing{i}_{\max}^{\texttt{small}})$}
		\STATE
		\STATE Compute $\var{e}_{\ing{k}}$ as in \eqref{eq:entropy}
		\ENDFOR
		\STATE $\ing{k}^{\star} := \argmax_{\ing{k}} \var{e}_{\ing{k}}$
		\STATE $\hat{\mtrx{w}}:=\mathcal{G}(
\Theta, \mtrx{M}, \{\mathcal{S}^{\text{PEW}}_{\mu}(\cdot|\Gamma_{\ing{k}^{\star}})\}_{\mu}, 
\mu_{\ing{k}^{\star}}, F_{\alpha}, \mtrx{Z}_{\ing{k}^{\star}},  \beta, \ing{i}_{\max}^{\texttt{large}}
).$
		\RETURN $\hat{\mtrx{w}}$
	\end{algorithmic}

\end{algorithm}

\subsubsection{Post-processing and overlap-add synthesis}
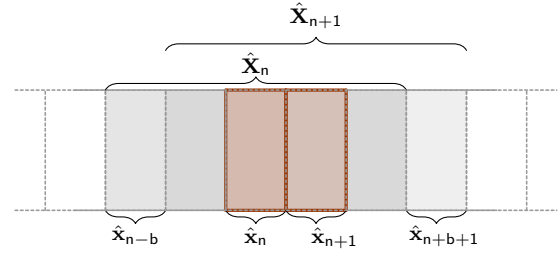
\begin{figure}[htb]
	\centering
	\begin{tikzpicture}[line cap=round,line join=round,>=triangle 45,x=0.8cm,y=0.4cm]
	
	\fill[line width=0.6pt,dash pattern=on 1pt off 1pt,color=lgray,fill=lgray,fill opacity=0.25] (0.,0.) -- (1.,0.) -- (1.,4.) -- (0.,4.) ;
	\fill[line width=0.6pt,dash pattern=on 1pt off 1pt,color=lgray,fill=lgray,fill opacity=0.25] (1.,0.) -- (2.,0.) -- (2.,4.) -- (1.,4.) ;
	\fill[line width=1.2pt,color=lred,fill=lred,fill opacity=0.3] (2.,0.) -- (3.,0.) -- (3.,4.) -- (2.,4.);
	\fill[line width=0.6pt,dash pattern=on 1pt off 1pt,color=lgray,fill=lgray,fill opacity=0.25] (3.,0.) -- (4.,0.) -- (4.,4.) -- (3.,4.);
	\fill[line width=0.6pt,dash pattern=on 1pt off 1pt,color=lgray,fill=lgray,fill opacity=0.25] (4.,0.) -- (5.,0.) -- (5.,4.) -- (4.,4.);
	\draw [line width=0.6pt,dash pattern=on 1pt off 1pt,color=lgray] (-1.5,0.)-- (6.5,0.);
	\draw [line width=0.6pt,dash pattern=on 1pt off 1pt,color=lgray] (1.,0.)-- (1.,4.);
	\draw [line width=0.6pt,dash pattern=on 1pt off 1pt,color=lgray] (-1.5,4.)-- (6.5,4.);
	\draw [line width=0.6pt,dash pattern=on 1pt off 1pt,color=lgray] (0.,0.)-- (0.,4.);
	
	\draw [line width=1.2pt,color=lred] (2.,0.)-- (3.,0.);
	\draw [line width=1.2pt,color=lred] (3.,0.)-- (3.,4.);
	\draw [line width=1.2pt,color=lred] (3.,4.)-- (2.,4.);
	\draw [line width=1.2pt,color=lred] (2.,4.)-- (2.,0.);
	
	\draw [line width=0.6pt,dash pattern=on 1pt off 1pt,color=lgray] (-1.,0.)-- (-1.,4.);
	\draw [line width=0.6pt,dash pattern=on 1pt off 1pt,color=lgray] (6.,0.)-- (6.,4.);
	\draw [line width=0.6pt,dash pattern=on 1pt off 1pt,color=lgray] (2.,4.)-- (2.,0.);
	\draw [line width=0.6pt,dash pattern=on 1pt off 1pt,color=lgray] (4.,0.)-- (4.,4.);
	\draw [line width=0.6pt,dash pattern=on 1pt off 1pt,color=lgray] (3.,4.)-- (3.,0.);
	\draw [line width=0.6pt,dash pattern=on 1pt off 1pt,color=lgray] (5.,0.)-- (5.,4.);
	
	\draw [black,decorate,decoration={brace,amplitude=5pt},
	xshift=0pt,yshift=0pt] (1,0) -- (0,0)
	node [black,midway,below=2pt,xshift=0pt] {\footnotesize $\hat{\vect{x}}_{\ing{n-b}}$};
	\draw [black,decorate,decoration={brace,amplitude=5pt},
	xshift=0pt,yshift=0pt] (3,0) -- (2,0)
	node [black,midway,below=3pt,xshift=0pt] {\footnotesize $\hat{\vect{x}}_{\ing{n}}$};
	\draw [black,decorate,decoration={brace,amplitude=5pt},
	xshift=0pt,yshift=0pt] (0,4) -- (5,4)
	node [black,midway,above=2pt,xshift=0pt] {\footnotesize $\hat{\mtrx{X}}_{\ing{n}}$};

	\begin{scope}[shift={(1,0)}]
	
	\fill[line width=0.6pt,dash pattern=on 1pt off 1pt,color=lgray,fill=lgray,fill opacity=0.15] (0.,0.) -- (1.,0.) -- (1.,4.) -- (0.,4.) ;
	\fill[line width=0.6pt,dash pattern=on 1pt off 1pt,color=lgray,fill=lgray,fill opacity=0.15] (1.,0.) -- (2.,0.) -- (2.,4.) -- (1.,4.) ;
	\fill[line width=1.2pt,color=lred,fill=lred,fill opacity=0.2] (2.,0.) -- (3.,0.) -- (3.,4.) -- (2.,4.);
	\fill[line width=0.6pt,dash pattern=on 1pt off 1pt,color=lgray,fill=lgray,fill opacity=0.15] (3.,0.) -- (4.,0.) -- (4.,4.) -- (3.,4.);
	\fill[line width=0.6pt,dash pattern=on 1pt off 1pt,color=lgray,fill=lgray,fill opacity=0.15] (4.,0.) -- (5.,0.) -- (5.,4.) -- (4.,4.);
	\draw [line width=1.2pt,color=lred] (2.,0.)-- (3.,0.);
	\draw [line width=1.2pt,color=lred] (3.,0.)-- (3.,4.);
	\draw [line width=1.2pt,color=lred] (3.,4.)-- (2.,4.);
	\draw [line width=1.2pt,color=lred] (2.,4.)-- (2.,0.);
	\draw [black,decorate,decoration={brace,amplitude=5pt},
	xshift=0pt,yshift=0pt] (3,0) -- (2,0)
	node [black,midway,below=3pt,xshift=7pt] {\footnotesize $\hat{\vect{x}}_{\ing{n+1}}$};
	\draw [black,decorate,decoration={brace,amplitude=5pt},
	xshift=0pt,yshift=15pt] (0,4) -- (5,4)
	node [black,midway,above=5pt,xshift=0pt] {\footnotesize $\hat{\mtrx{X}}_{\ing{n+1}}$};
	\draw [black,decorate,decoration={brace,amplitude=5pt},
	xshift=0pt,yshift=0pt] (5,0) -- (4,0)
	node [black,midway,below=2pt,xshift=3pt] {\footnotesize $\hat{\vect{x}}_{\ing{n+b+1}}$};
	
	\draw [line width=0.6pt,dash pattern=on 1pt off 1pt,color=lgray] (-1.5,0.)-- (6.5,0.);
	\draw [line width=0.6pt,dash pattern=on 1pt off 1pt,color=lgray] (1.,0.)-- (1.,4.);
	\draw [line width=0.6pt,dash pattern=on 1pt off 1pt,color=lgray] (-1.5,4.)-- (6.5,4.);
	\draw [line width=0.6pt,dash pattern=on 1pt off 1pt,color=lgray] (0.,0.)-- (0.,4.);

	\draw [line width=0.6pt,dash pattern=on 1pt off 1pt,color=lgray] (-1.,0.)-- (-1.,4.);
	\draw [line width=0.6pt,dash pattern=on 1pt off 1pt,color=lgray] (6.,0.)-- (6.,4.);
	\draw [line width=0.6pt,dash pattern=on 1pt off 1pt,color=lgray] (2.,4.)-- (2.,0.);
	\draw [line width=0.6pt,dash pattern=on 1pt off 1pt,color=lgray] (4.,0.)-- (4.,4.);
	\draw [line width=0.6pt,dash pattern=on 1pt off 1pt,color=lgray] (3.,4.)-- (3.,0.);
	\draw [line width=0.6pt,dash pattern=on 1pt off 1pt,color=lgray] (5.,0.)-- (5.,4.);

	\end{scope}
	
	\end{tikzpicture}
	
	\caption{Segment processing for frame $\ing{n}$ and frame $\ing{n+1}$.	\label{fig:framessketch}
}
	
\end{figure}
We recall that the denoiser is applied in an frame-based scenario: given noisy frame(s) $\mtrx{Y}_{\ing{n}}$, the denoisers output estimated frame(s) $\hat{\mtrx{X}}_{\ing{n}}$ which need to be transformed back into a full time-domain signal $\vect{x}$. For this, we first need to extract from $\hat{\mtrx{X}}_{\ing{n}}$ a single estimated frame $\hat{\vect{x}}_{\ing{n}}$:
\begin{itemize}
\item in the plain sparse case, this is straightforward as $\hat{\mtrx{X}}_{\ing{n}} \Rset{L}{1}$ is already a vector; 
\item for the social sparse case, we set $\hat{\vect{x}}_{\ing{n}} = \hat{\mtrx{X}}_{\ing{n}}(:,\ing{b}+1)$ to be the central column of the matrix $\hat{\mtrx{X}}_{\ing{n}}$, see Fig.~\ref{fig:framessketch}. 
\end{itemize}

Given the estimated frames $\{\hat{\vect{x}}_{\ing{n}}\}_{\ing{n}}$, and before the final overlap-add that will lead to the full time-domain estimate $\hat{\vect{x}}$, we perform a simple frequency-domain Wiener filtering on each $\hat{\vect{x}}_{\ing{n}}$ similar to the one used in the Block-Thresholding algorithm \cite{yu2008audio} which we will use as a comparison in Section~\ref{sec:expe_denoising}.
Such a Wiener filtering requires an estimation of the noise power $\sigma^2$, as well as an estimation of the signal power, both in the frequency domain. For the latter, we use the squared magnitudes of $\mtrx{A}\hat{\vect{x}}$ (resp. of $\htransp{\mtrx{D}}\hat{\vect{x}}$). Oracle values of $\sigma^{2}$ will be used in the experiments. Practically, we observed that this post-processing is useful at very low SNR (\emph{i.e} $0 \text{ dB}$) where we observe ``musical noise'' effect. 

Finally, overlap-add synthesis is performed, taking into account the windows that were applied onto the frames to get the noisy frames $\mtrx{Y}_{\ing{n}}$.

\subsection{Experimental Study\label{sec:expe_denoising}}

This section aims at comparing effects of the different shrinkages (plain or social), the different models (synthesis or analysis), and the degradation level on the audio denoising performance. 

\paragraph{Datasets}
We conduct experiments on excerpts from the RWC Music Database \cite{goto2002RWC}. We use the ``Pop'' and ``Jazz'' genres as is, and subcategorize the ``Classic'' genre (Vocals, Chamber, Symphonies), leading to 5 subsets. All the tracks are sufficiently diverse to reflect the robustness of the approach on different audio content. For each subset of the database, we contaminated an excerpt of each available track in order to get around one hour of noisy material for each category. We also perform experiments on the TIMIT database \cite{garofolo1993darpa} for evaluation on speech content. All the audio examples used here are also down-sampled to 16 kHz.\\

\paragraph{Performance measures}
We use as a first objective recovery performance numerical measure the Signal-to-Noise Ratio (SNR) difference between the noisy and enhanced signals. We also compare the perceptual speech quality with the objective MOS-PESQ scores~\cite{rix2001perceptual}, and analyse the intelligibility through the Short-Time Objective Intelligibility index (STOI)~\cite{taal2011algorithm}. For music, we compare the perceptual quality on the entire RWC excerpts collection with the Objective Difference Grade (ODG) PEAQ scores~\cite{kabal2002peaq}. Finally, for all methods we compare computation times in seconds.\\

\paragraph{Compared methods}
We consider the plain sparse, plain cosparse, social sparse and social cosparse denoisers, as well as the state-of-the-art time-frequency block thresholding (BT)~\cite{yu2008audio}. The main parameters are set as follows:
\begin{itemize}
	\item frame size: $\ing{L} = 64$ ms for music $\ing{L} = 32$ ms for speech;
	\item overlap, $ 75\%$;
	\item number of overlapping segments for social denoisers: $\ing{b} = 5$ for music, $\ing{b} = 1$ for speech;
	\item time-frequency patterns for social denoisers: $\left\{ \Gamma_{\ing{k}}\right\}_{\ing{k}=1}^{\ing{K}}$ presented on Fig.~\ref{fig:Stencils} for music and Fig.~\ref{fig:SpeechStencils} for speech.
	\item stopping criteria $ \beta = 10^{-3}$, $\ing{i}_{\max}^{\texttt{small}}=10$, $\ing{i}_{\max}^{\texttt{large}}=10^{6}$;
\end{itemize}
The time-frequency synthesis/analysis operators are:
\begin{itemize}	
	\item {\bf Synthesis operator}: $\mtrx{D}$ is the inverse DFT of redundancy $\ing{R}$, that is to say $\mtrx{D}\Cset{L}{S}$ with $\ing{S}=(\ing{R}\times\ing{L})$ and $\mtrx{D}_{\ing{l}\ing{s}} := \ing{S}^{-1/2}e^{j\frac{2\pi\ing{l}\ing{s}}{\ing{S}}}$. One can check that $\mtrx{D}\htransp{\mtrx{D}}=\mtrx{I}$;
	\item {\bf Analysis operator}: $\mtrx{A}$ is the forward DFT of redundancy $\ing{R}$, that is to say $\mtrx{A}\Cset{P}{L}$ with $\ing{P}=(\ing{R}\times\ing{L})$ and $\mtrx{A}_{\ing{p}\ing{l}} = \ing{P}^{-1/2}e^{-j\frac{2\pi\ing{p}\ing{l}}{\ing{P}}}$. Again, one can check that $\htransp{\mtrx{A}}\mtrx{A}=\mtrx{I}$.
\end{itemize}
Practically, products with $\mtrx{A}$ (resp. $\htransp{\mtrx{D}}$), are done using the FFT of size $\ing{P}$ (resp. $\ing{S}$) on a zero-padded signal of initial length $\ing{L}$. Similarly, products with $\htransp{\mtrx{A}}$ (resp. $\mtrx{D}$), are done by truncating the inverse fast transform.

All denoisers require a parameter $\varepsilon$  ruling the $l_2$ regularization for the denoising constraint.
For the plain sparse denoisers, $\varepsilon$ is set to $\sigma \sqrt{\sum_{\ing{j}=1}^{\ing{L}} \vect{w}_{\ing{j}} } $, with $\vect{w}_{\ing{j}}$ the j\textsuperscript{th} entry of the window $\vect{w}$ and $\sigma^2$ the known noise variance. 
For the adaptive social sparse denoisers, we scale $\varepsilon$ to $(2\ing{b}+1) \sigma \sqrt{\sum_{\ing{j}=1}^{\ing{L}} \vect{w}_{\ing{j}} }$.

The adaptive social sparse denoisers also require to set $\iter{\mu_{\ing{k}}}{0}$ and $\alpha$ (see Algorithm~\ref{alg:AUDASCITY}). To adapt these parameters to the local peak audio level $\norm{ \text{vec}(\mtrx{Y})}{\infty}$ and to the number of active bins in the time-frequency pattern $\Gamma_{\ing{k}}$, we set 
\begin{eqnarray}
\iter{\mu_{\ing{k}}}{0} &:=& \norm{\Gamma_{\ing{k}}}{0} \times \norm{ \text{vec}(\mtrx{Y})}{\infty}\\
\alpha &:=& \min \left( \frac{\sigma}{\sqrt{\text{var}(\text{vec}(\mtrx{Y}))} }, 0.99 \right),
\end{eqnarray}
where $\text{vec}(\cdot)$ vectorizes the matrix. This parameterization reflects the ``instantaneous'' SNR in the region being processed. The two parameters $\alpha$ and $\mu$ rule how aggressively the sparse regularization is performed. 

\paragraph{Pilot study}

Given the large combinatorics of experiments related to all possible configurations (plain/social, analysis/synthesis, redundancy factor) and noise levels, we performed a first pilot study for two input noise levels (5~dB, 20~dB). Each configuration was tested over five 10~second excerpts  from the RWC database covering the five music genre subsets. The average SNR improvements, as well the average computation times\footnote{All reported computation times were measured using a \Matlab implementation of the algorithms on a laptop equipped with a 2.8 Ghz \IntelProc processor and 16 GB of RAM memory.} (relative to the audio duration) are summarized in Table~\ref{tab:DenoisingPilotStudy}. 
\begin{table}[ht]
	\centering
	\caption{Pilot study: SNR improvements ($\Delta$SNR) and processing times\protect\\ relative to audio duration ($\times$RT) for all configurations\label{tab:DenoisingPilotStudy}}
	\subfloat[Input SNR: 5~dB\label{tab:DenPilot5dB}]{
\resizebox{0.48\textwidth}{!}{\begin{tabular}{l|c|c|c|c|c|c|c|c|}
	\cline{2-9}
	& \multicolumn{4}{c|}{Plain} & \multicolumn{4}{c|}{Social} \\ \cline{2-9} 
	& \multicolumn{2}{c|}{Analysis} & \multicolumn{2}{c|}{Synthesis} & \multicolumn{2}{c|}{Analysis} & \multicolumn{2}{c|}{Synthesis} \\ \cline{2-9} 
	& $\Delta$SNR & $\times$RT & $\Delta$SNR & $\times$RT & $\Delta$SNR & $\times$RT & $\Delta$SNR & $\times$RT \\ \hline
	\multicolumn{1}{|l|}{$\ing{R}=1$} & 7.04 & 2.9  & 7.04 & 4.0 & 7.70 & 5.9 & 7.70 & 8.1 \\ \hline
	\multicolumn{1}{|l|}{$\ing{R}=2$} & 7.29 & 10.0 & 7.30 & 13.3 & 7.75 & 12.8 & 7.73 & 16.6 \\ \hline
	\multicolumn{1}{|l|}{$\ing{R}=4$} & 7.26 & 20.9 & 7.26 & 26.3 & 7.78 & 24.5 & 7.74 & 31.5 \\ \hline
\end{tabular}}}\\

	\subfloat[Input SNR: 20~dB\label{tab:DenPilot20dB}]{
\resizebox{0.48\textwidth}{!}{\begin{tabular}{l|c|c|c|c|c|c|c|c|}
	\cline{2-9}
	& \multicolumn{4}{c|}{Plain} & \multicolumn{4}{c|}{Social} \\ \cline{2-9} 
	& \multicolumn{2}{c|}{Analysis} & \multicolumn{2}{c|}{Synthesis} & \multicolumn{2}{c|}{Analysis} & \multicolumn{2}{c|}{Synthesis} \\ \cline{2-9} 
	& $\Delta$SNR & $\times$RT & $\Delta$SNR & $\times$RT & $\Delta$SNR & $\times$RT & $\Delta$SNR & $\times$RT \\ \hline
	\multicolumn{1}{|l|}{$\ing{R}=1$} & 3.17 & 5.9 & 3.17 & 8.0 & 3.21 & 2.2 & 3.21 & 2.9 \\ \hline
	\multicolumn{1}{|l|}{$\ing{R}=2$} & 3.32 & 13.9 & 3.31 & 17.4 & 3.18 & 5.0 & 3.19 & 6.5 \\ \hline
	\multicolumn{1}{|l|}{$\ing{R}=4$} & 3.29 & 26.4 & 3.29 & 34.1 & 3.24 & 10.9 & 3.23 & 13.8 \\ \hline
\end{tabular}}}

\end{table}

We observe that:
\begin{itemize}
\item For each noise level, each redundancy, and each thresholding operator, the performance of the analysis and synthesis models in decibels is almost identical, while the analysis version is by $20\%$ to nearly $40\%$ faster that the synthesis version. As a result the rest of the experiments are conducted only with the analysis version.
\item All other factors being equal, the computation time is roughly proportional to the redundancy $\ing{R}$, while the improvement in ouput SNR is often very limited. In the rest of the experiments we thus choose $\ing{R}=2$, which seems to give the best compromise (and in fact, even the best performance in many configurations). It also enables a transparent comparison with the baseline defined by block thresholding.
\end{itemize}

\paragraph{Denoising performance}

\begin{figure*}[!htbp]
	\centering
	\subfloat[RWC Jazz]{\includegraphics[width=5cm]{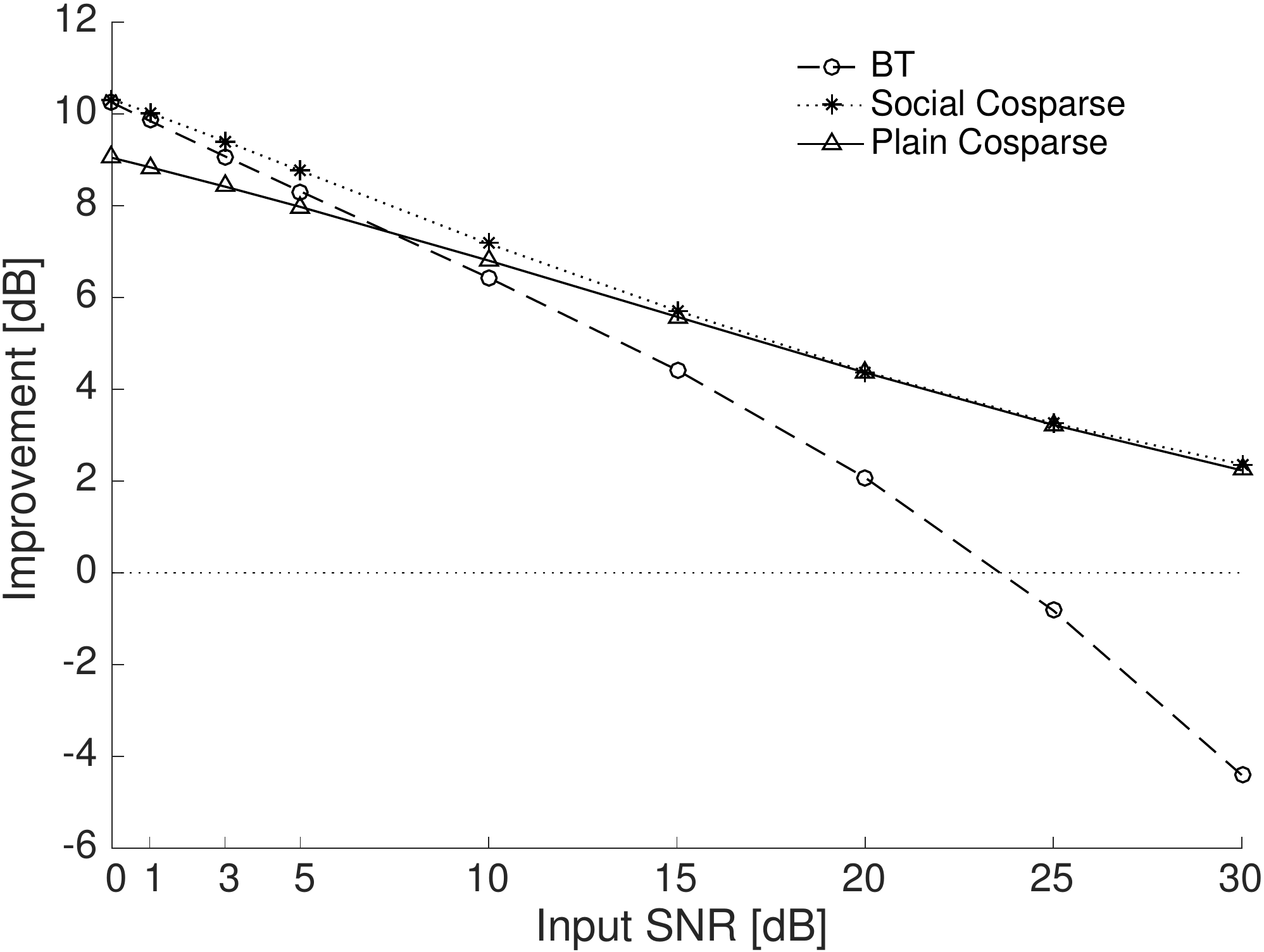}%
		\label{fig:JazzDen}}
	\hfil
	\subfloat[RWC Pop]{\includegraphics[width=5cm]{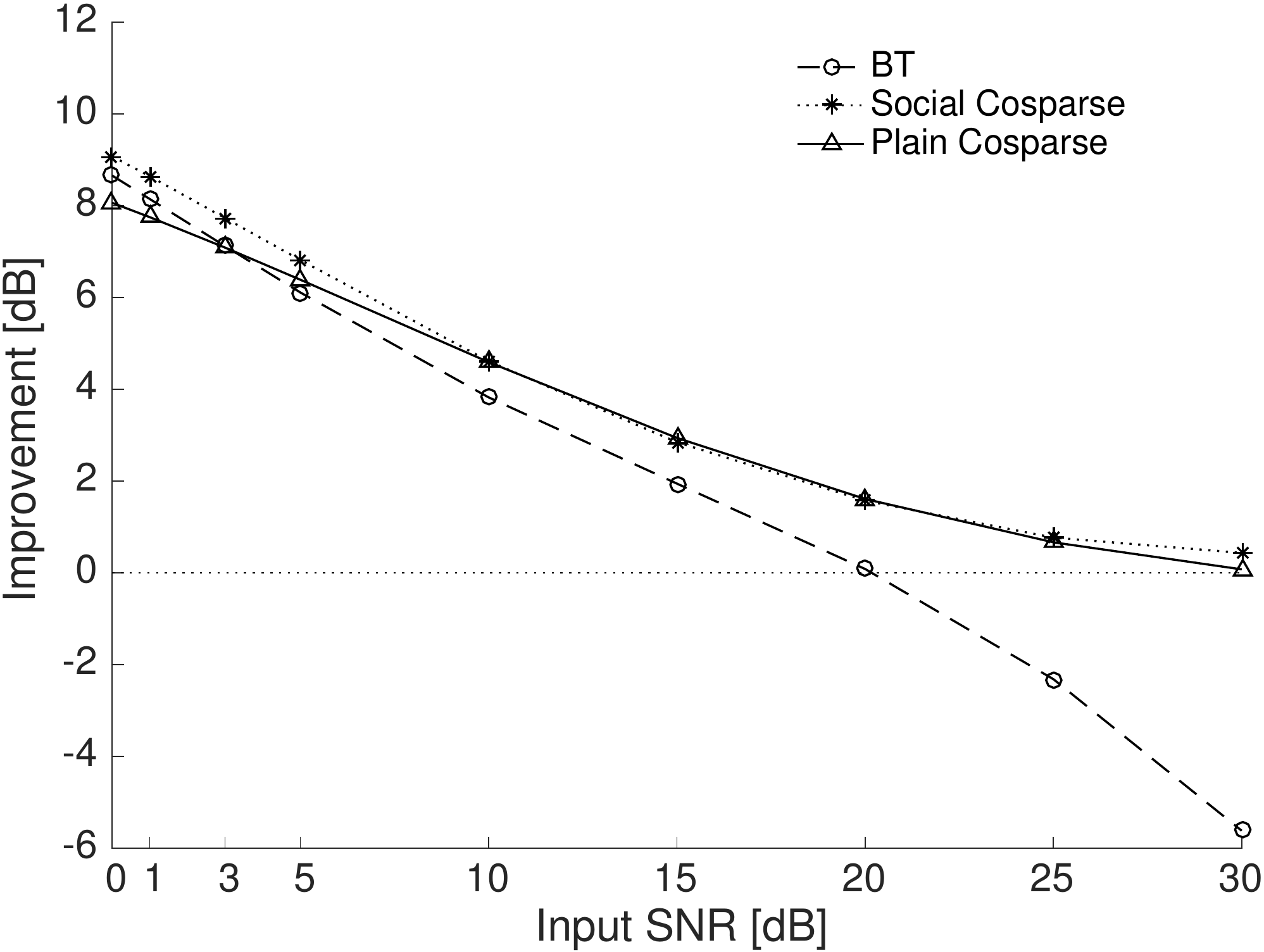}%
		\label{fig:PopDen}}
	\hfil
	\subfloat[RWC Classic: Chamber]{\includegraphics[width=5cm]{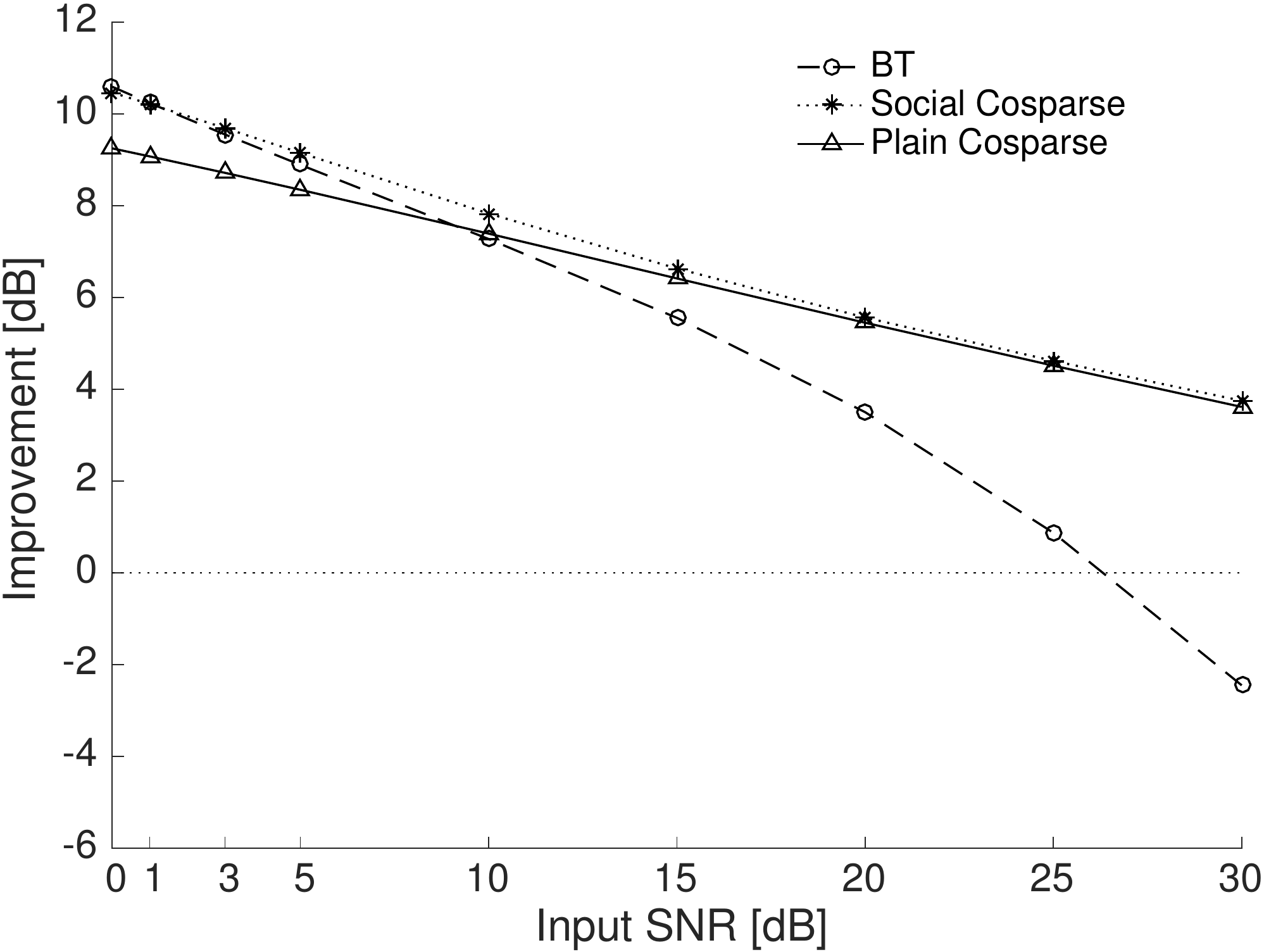}%
		\label{fig:ChamberDen}}
	\hfil
	\subfloat[RWC Classic: Vocals]{\includegraphics[width=5cm]{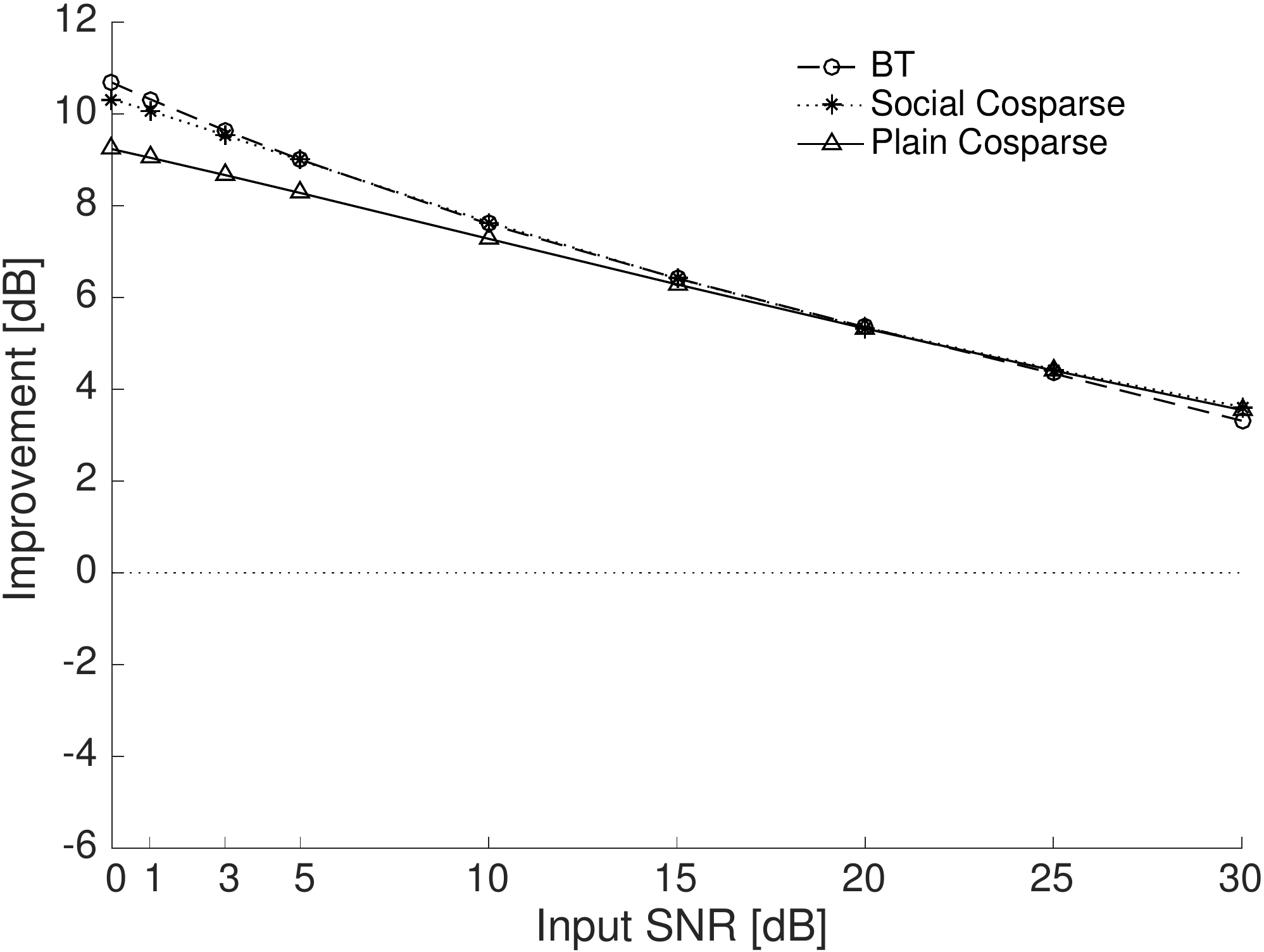}%
		\label{fig:VocalsDen}}
	\hfil
	\subfloat[RWC Classic: Symphonies]{\includegraphics[width=5cm]{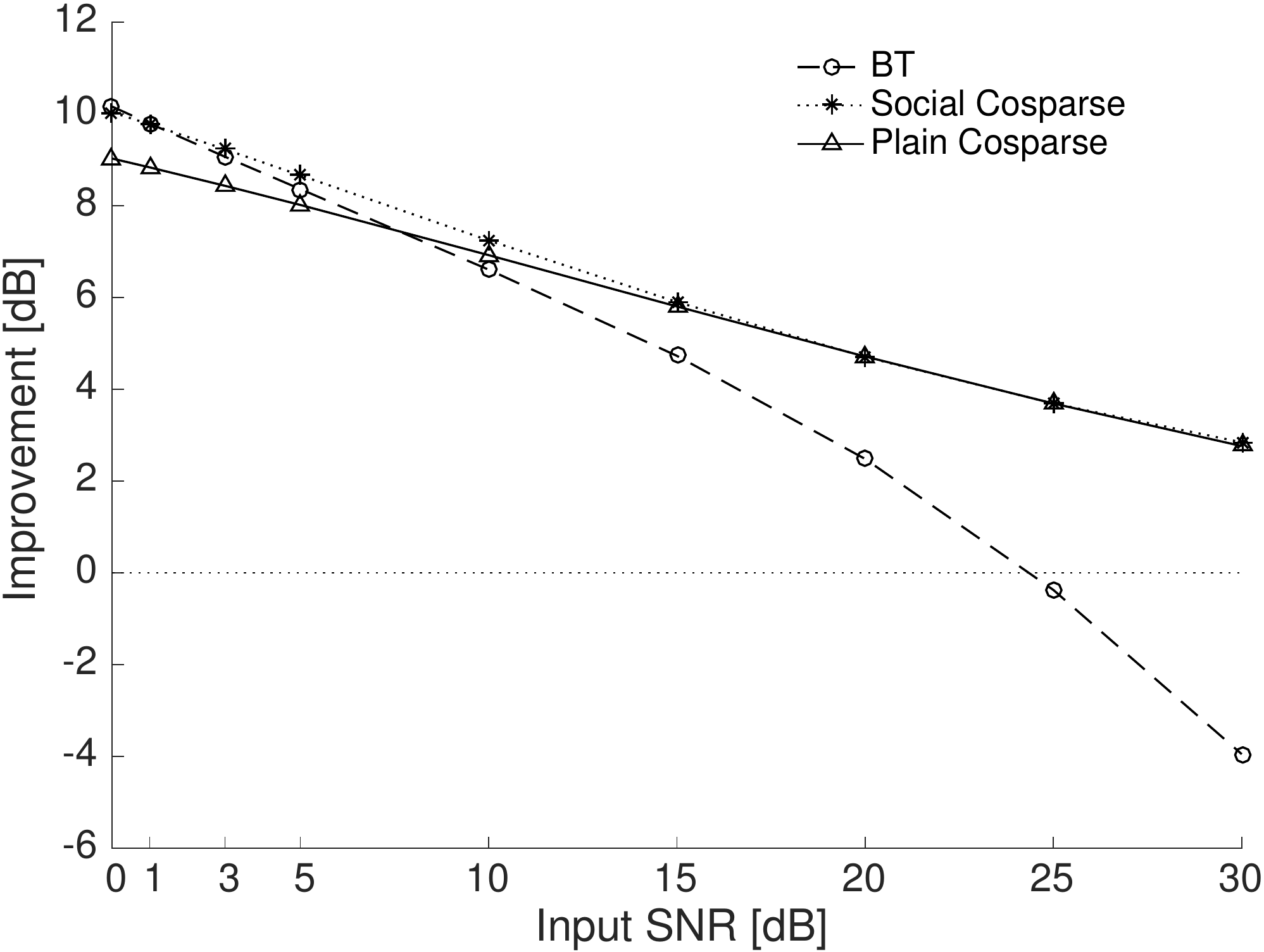}%
		\label{fig:SymphoniesDen}}
	\hfil
		\subfloat[TIMIT]{\includegraphics[width=5cm]{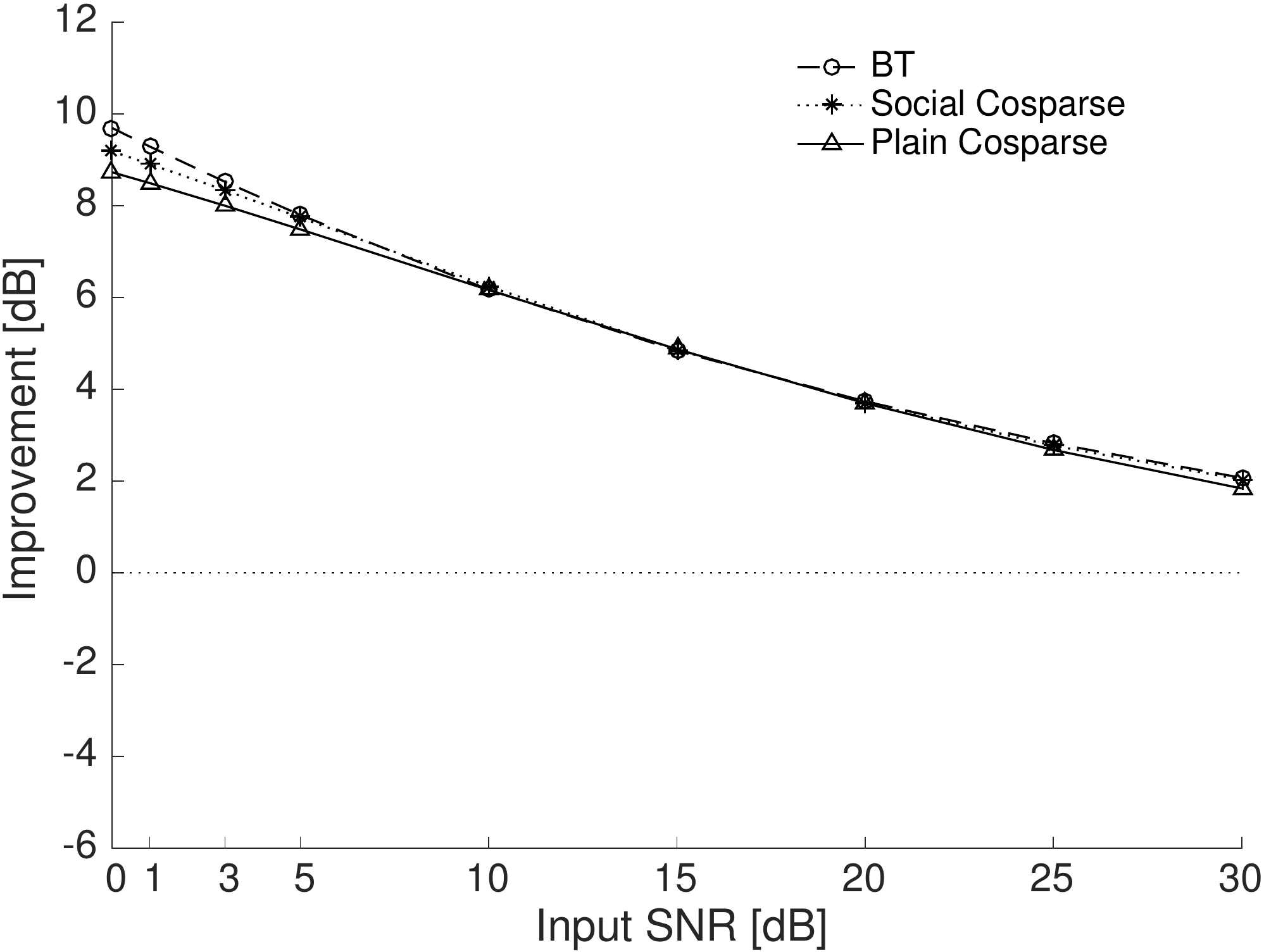}%
		\label{fig:SpeechDen}}
	\hfil
\begin{center}	{\centering\caption{Denoising Numerical Results: SNR improvement {[}dB{]	\label{fig:ResNumNoise}
}}}
\end{center}
\end{figure*}

Given the pilot study, we now focus on the cosparse denoisers (plain and social) as well as block thresholding, all with redundancy $\ing{R}=2$. We consider nine input SNR levels in dB: \{$0$, $1$, $3$, $5$, $10$, $15$, $20$, $25$, $30$\} and work with the full datasets. 

Figure \ref{fig:ResNumNoise} shows averaged SNR improvements over each of the 5 music subsets as well as the TIMIT speech dataset.

Results on Fig.~\ref{fig:ResNumNoise} show that either the social cosparse or the plain cosparse algorithms outperform Block Thresholding (BT) on the denoising task for almost every category of audio content. The benefit given by the social flavour of the algorithm is widely seen at low SNRs where the social method and BT have comparable performance. Indeed, the social version perform 1 to 3 \text{dB} better than the simple cosparse version for input SNRs below 10. These results strengthen the idea that adaptiveness can be beneficial for highly degraded conditions. Moreover, the simple sparse approach catch up with or even overstep the social one for higher SNRs. This way, we can guess that recovering hidden structure is optional as the signal is already well clustered in the time-frequency plane for light noise conditions. The difference between our approaches and BT increases with the input SNR. We note a significant contrast at high SNR where BT underperforms by more than 6 dB in the less favorable configuration. This might be because BT strongly relies on the noise model whereas cosparse and social cosparse methods try to emphasize the signal itself.

We gathered standard deviation informations associated to Figure~\ref{fig:ResNumNoise} and results demonstrate that the plain cosparse denoiser produces less variable results as the standard deviation is the lowest for this technique in 80\% of the tested cases. We also notice that, without considering any specific algorithm, the improvement variability seems to increase with the input SNR. Indeed, for light noise conditions, the standard deviation reaches up to 3.29 dB for BT on the RWC ``Solo'' musical excerpts.

\begin{figure*}[!htbp]
	\centering
	\subfloat[TIMIT: STOI Index]{\includegraphics[width=5cm]{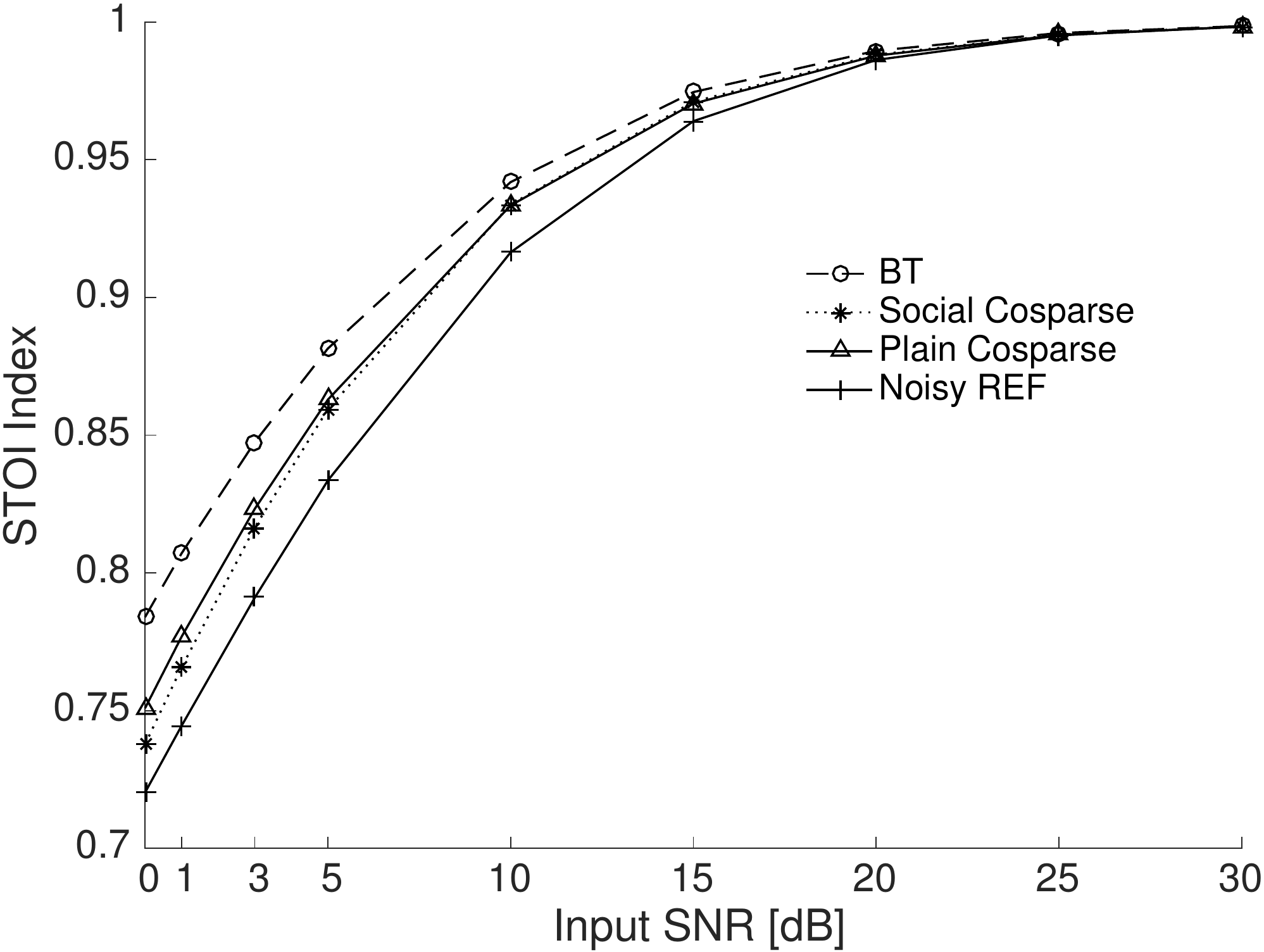}%
		\label{fig:STOITimit}}
	\hfil
	\subfloat[TIMIT: MOS-Mapped PESQ Value]{\includegraphics[width=5cm]{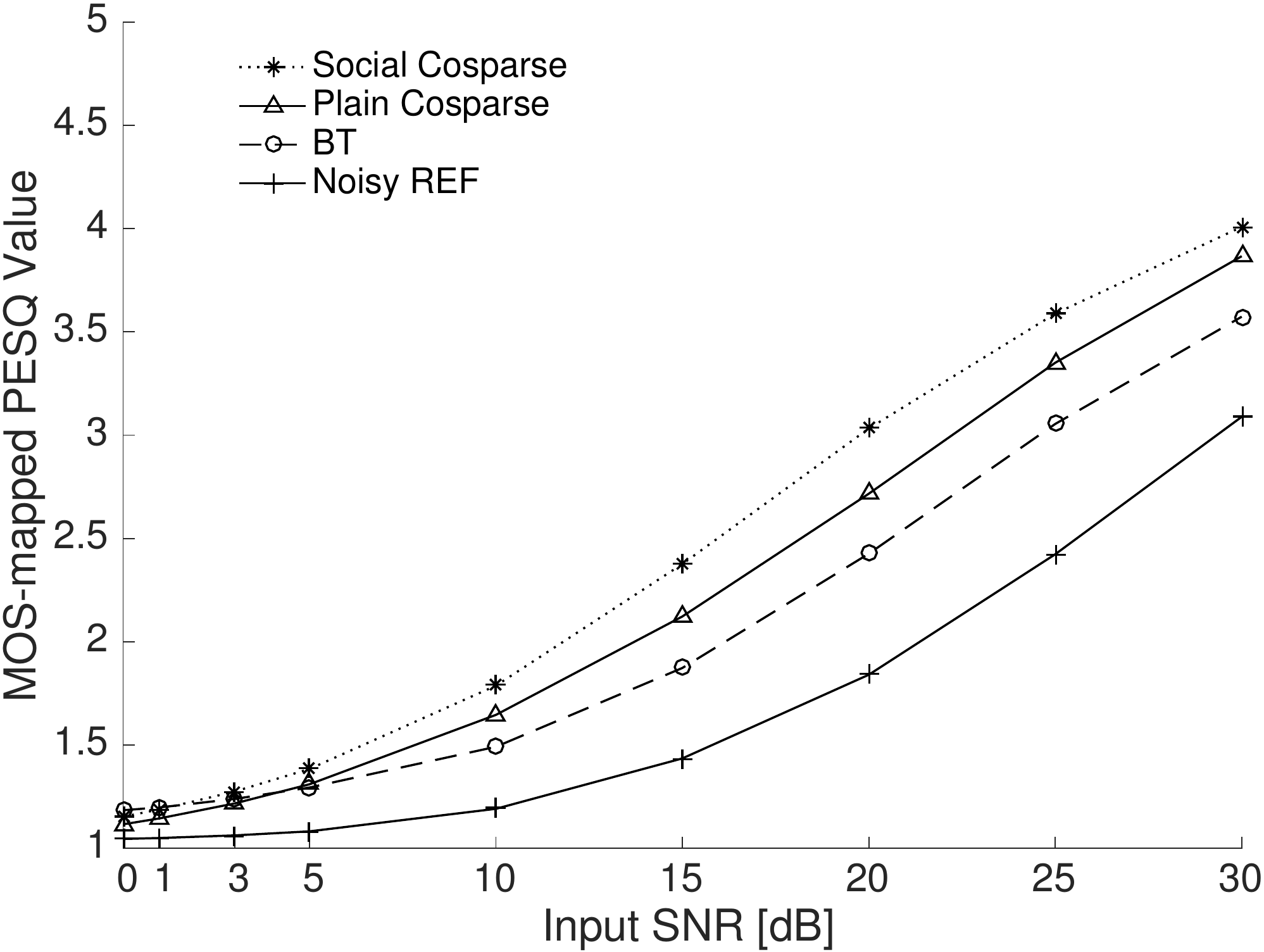}%
		\label{fig:PESQTimit}}
	\hfil
	\subfloat[RWC: PEAQ ODG score]{\includegraphics[width=5cm]{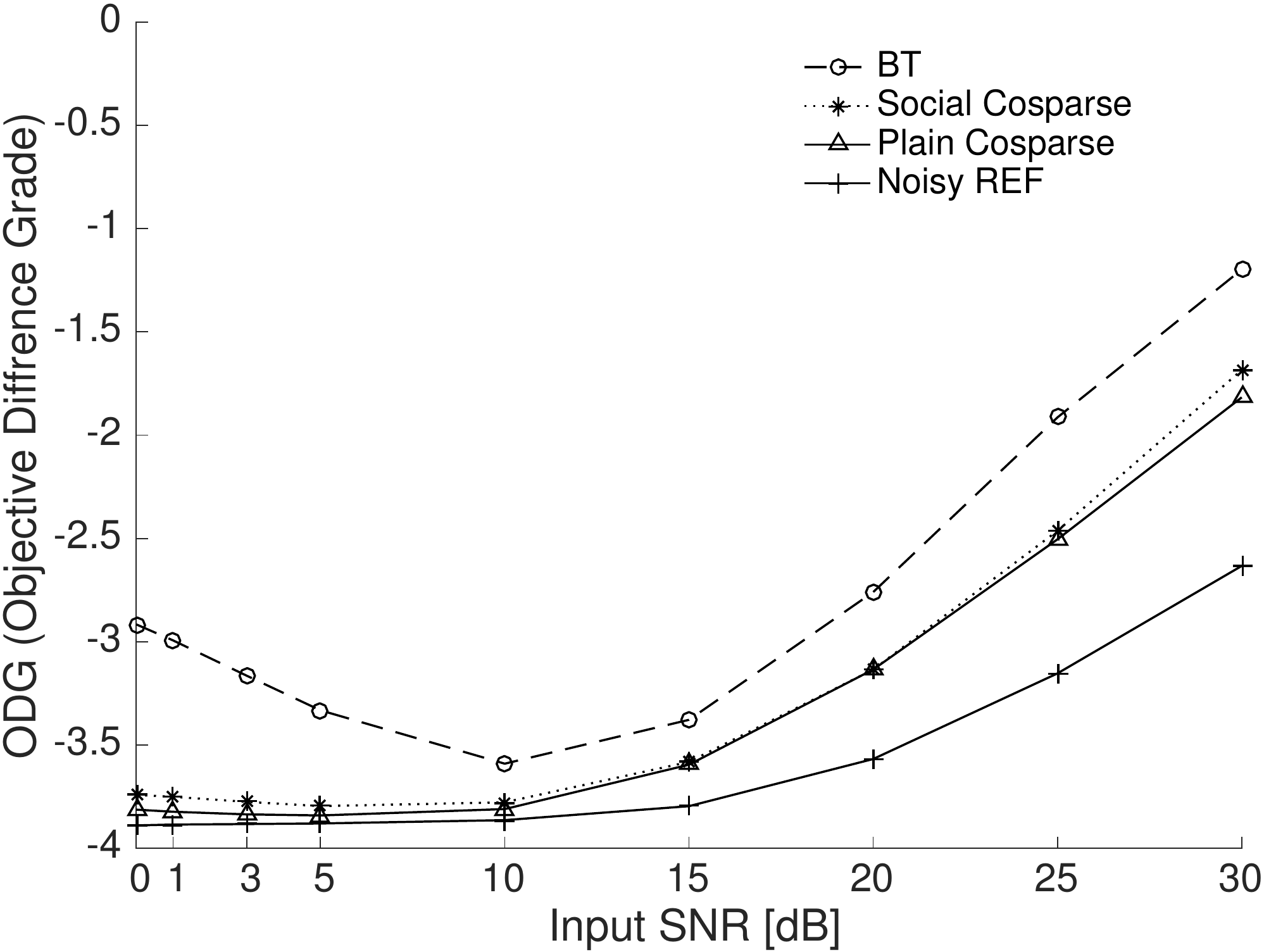}}
		\label{fig:PEAQDenRWC}
	\hfil
	\caption{Denoising Objective Quality and Intelligibility Results}
	\label{fig:ResObjSpeechNoise}
\end{figure*}

Figure \ref{fig:ResObjSpeechNoise} shows averaged STOI/PESQ/PEAQ performance over each of the 5 music subsets as well as the TIMIT speech dataset. Even if Fig.~\ref{fig:VocalsDen} and~\ref{fig:SpeechDen} do not show clear superiority of one or another method on SNR improvement for voice based audio content, Fig.~\ref{fig:PESQTimit} reveals improved objective speech quality (PESQ metric) for both social and plain cosparse denoisers. 

\paragraph{Computation time\label{par:CompTimeDen}}

\begin{table}
	\centering
	\caption{Computational performance of cosparse denoisers \label{tab:DenComp}}
	\begin{tabular}{|l||c|c|c|c|}
		\hline
		\multirow{2}{*}{\begin{tabular}[c]{@{}l@{}}Input\\ SNR {[}dB{]}\end{tabular}} & \multicolumn{2}{c|}{Plain cosparse}                                                                                              & \multicolumn{2}{c|}{Social cosparse}                                                                                       \\ \cline{2-5} 
		&  \begin{tabular}[c]{@{}c@{}}$\Delta$ SNR\end{tabular} & \begin{tabular}[c]{@{}c@{}}$\times$ RT\end{tabular} & \begin{tabular}[c]{@{}c@{}}$\Delta$ SNR\end{tabular} & \begin{tabular}[c]{@{}c@{}}$\times$ RT\end{tabular} \\ \hline
		0 & 9.45 & 7.8  & 9.50  & 16.6 \\ \hline
		5 & 7.62 & 10.3 & 7.76  & 12.0 \\ \hline
		10 & 5.91  & 12.6  & 6.03  & 9.2 \\ \hline
		15 & 4.33  & 16.6  & 4.50  & 7.0 \\ \hline
		20 & 3.02  & 22.1  & 3.18  & 5.5 \\ \hline
	\end{tabular}
\end{table}

For the social case, the computational cost is driven by the shrinkage (PEW) and the projection steps. However, evaluating PEW shrinkage is relatively fast, as it can be computed through 2-D convolution in the time-frequency domain. Besides, since we set low $\ing{i}_{\max}$ for the initialization loop, the choice of $\Gamma$ is quite fast and adds only $(\ing{b}-1)\times\ing{i}_{\max}^{\texttt{small}}$ iterations compared to the case where only one time-frequency pattern is considered. These properties allow to expect the social cosparse denoiser to have runtime comparable to that of the plain cosparse denoiser.

Table~\ref{tab:DenComp} displays processing times in seconds for both denoising procedures. These computational comparisons are conducted  on a single 10 second sound excerpt for each genre. The SNR improvements are in line with what was previously observed on larger datasets with very similar performance of both methods. However we note very different behaviors between the two cosparse denoisers in terms of runtime. While the plain flavor is fastest at low SNR, the social version is fastest at high SNR. This suggests that in practice the choice of one of these methods might be rather driven by speed considerations than by denoising  performance.

\section{Audio Declipping Use Case}\label{sec:Declipping}

With denoising, declipping is another well-known problem in signal processing. Magnitude saturation can occur at different steps in the acquisition, reproduction or analog-to-digital conversion process. Restoring saturated signal is of great interest for many applications in digital communications, image processing or audio. In the latter, while light to moderate clipping cause only some audible clicks and pops, more severe saturation highly affect original signals which sound contaminated by rattle noise. The perceived degradation depends on the clipping level and the original signal. More recently, studies~\cite{tachioka2014speech,harvilla2014least} showed the negative impact of clipped signals when used in signal-processing pipelines leading to recognition, transcription or classification applications.  

\begin{figure}[!htbp]
	\centering
	\subfloat[Signal waveforms]{\includegraphics[width=7cm]{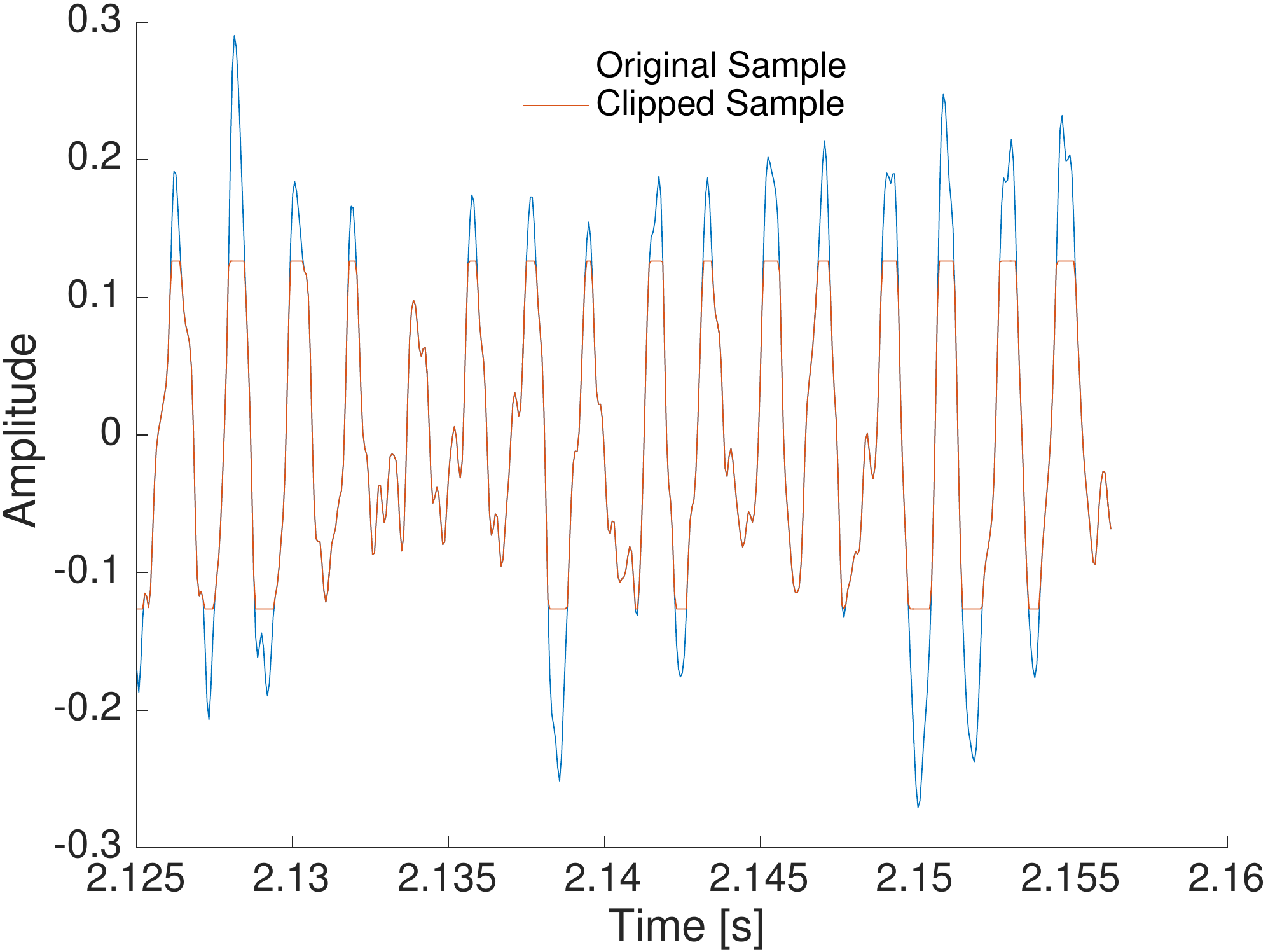}%
		\label{fig:ClippedWF}}
	\caption{Clipped signal example\label{fig:ClipSig}}

\end{figure}

In the following section, we use the idealized hard-clipping model below. Although simple, it correctly approximates the magnitude saturation and allows to split up the samples into a clipped set and a reliable set:
\begin{equation}
\label{eq:ClipModel}
\vect{y}_{\ing{i}\ing{j}} = \left\{ 
\begin{array}{l l}
\vect{x}_{\ing{i}\ing{j}} & \quad \text{for } \abs{\vect{x}_{\ing{i}\ing{j}}} \leq \tau;\\
\sign(\vect{x}_{\ing{i}\ing{j}})\tau & \quad \text{otherwise;}\\ \end{array} \right.
\end{equation}
with $\vect{y}_{\ing{i}\ing{j}}$ (resp. $\vect{x}_{\ing{i}\ing{j}}$) a sample from $\mtrx{Y}_{\ing{n}}$ (resp. $\mtrx{X}_{\ing{n}}$) and $\tau$ the hard-clipping level. A visual example of such a degradation is given on figure~\ref{fig:ClipSig}.
In real settings where softer saturation occurs, this model can be enforced with appropriate data pre-processing. 

\subsection{Generalized projections for the declipping problem}\label{subec:SolveDeclip}
Denote $\Omega_+$ (resp. $\Omega_-$) the collection of indices $\ing{ij}$ of the samples in matrix $\mtrx{Y}$ affected by positive (resp. negative) magnitude clipping. Similarly denote $\Omega_r$ the indices of the reliable samples (not affected by clipping), and for any of these sets $\Omega$ define $\mtrx{V}_{\Omega}$ the matrix formed by keeping only the entries of $\mtrx{V}$ indexed by $\Omega$ and setting the rest to zero. 
 
 The data-fidelity constraint can now be expressed
  for the analysis setting with $\mtrx{M}:=\mtrx{A}$ by $$\Theta := \left\{ 
\mtrx{W}\ |
\begin{array}{l}
\mtrx{W}_{\Omega_r} = \mtrx{Y}_{\Omega_r};\\
\mtrx{W}_{\Omega_+} \succcurlyeq \mtrx{Y}_{\Omega_+}; \\ 
\mtrx{W}_{\Omega_-} \preccurlyeq \mtrx{Y}_{\Omega_-}.\end{array} \right\}
$$
while for the synthesis setting, with $\mtrx{M}:=\mtrx{I}$, we set
 $$\Theta := \left\{ 
\mtrx{W}\ |
\begin{array}{l}
(\mtrx{D}\mtrx{W})_{\Omega_r} = \mtrx{Y}_{\Omega_r};\\
(\mtrx{D}\mtrx{W})_{\Omega_+} \succcurlyeq \mtrx{Y}_{\Omega_+}; \\ 
(\mtrx{D}\mtrx{W})_{\Omega_-} \preccurlyeq \mtrx{Y}_{\Omega_-}.\end{array} \right\}.$$
Similarly to the denoising use-case, these choices hold for both plain and social versions.

In the analysis setting, the desired projection reduces to component wise magnitude constraints (see. Appendix~\ref{app:ProxDeclipping}) and can be expressed as:  
$$[\mathcal{P}_{\Theta,\mtrx{M}}(\mtrx{Z})]_{(\ing{ij})}~=~\left\{ 
\begin{array}{l l}
\mtrx{Y}_{(\ing{ij})} & \text{if } \ing{ij}\in\Omega_r;\\
(\htransp{\mtrx{M}}\mtrx{Z})_{(\ing{ij})} & \text{if~}\left\{\begin{array}{l}\ing{ij}\in\Omega_+,(\htransp{\mtrx{M}}\mtrx{Z})_{(\ing{ij})} \geq \tau;\\ \text{or}\\\ing{ij}\in\Omega_-,(\htransp{\mtrx{M}}\mtrx{Z})_{(\ing{ij})} \leq -\tau;\end{array}\right.\\
\sign(\mtrx{Y}_{(\ing{ij})})\tau & \text{otherwise.}\\
\end{array} \right.$$

In this case, matrix-vector products with $\htransp{\mtrx{M}}$ dominates the computing cost of the generalized projection. When this can be done with a fast transform, the analysis flavor has once again low complexity.

For the synthesis case, the projection step is not that straightforward and needs to be computed with another nested iterative procedure. Even if exploiting the tight-frame property on $\mtrx{D}$ can help building an efficient algorithm for the projection, the overall computation cost for the synthesis flavor however remains substantially higher than the analysis version. To get more details on this projection, refer to Appendix~\ref{app:ProxDeclipping}.

\subsection{Algorithms for the declipping inverse problem\label{subsec:AlgoDeclip}}

With all the steps defined, we can now instantiate the general algorithm $\mathcal{G}$ in the different cases.

\subsubsection{Plain sparse audio declippers}\label{sec:SparseDeclippers}

Similarly to denoising, for both the analysis and the synthesis version, we instantiate the general algorithm $\mathcal{G}$ by choosing the operators described in Table~\ref{tab:SparseDeclip}.

The update rule $F$ for $\mu$ is set to gradually decrease $\mu$ by $1$ at each iteration, starting from $\iter{\mu}{0} = \ing{P} - 1$ for the analysis case (resp. $\iter{\mu}{0} = \ing{S} - 1$ for the synthesis case). This way, we relax the sparse constraint the same way we do it for denoising.

\begin{table}[h!]
	\caption{Parameters of Algorithm~\ref{alg:AbstractAlgo} for the Plain Sparse Declipper \label{tab:SparseDeclip}}
	\centering
		\begin{tabular}{m{0.45\columnwidth}|m{0.55\columnwidth}}
		
			\multicolumn{1}{c|}{\textbf{Analysis}}                                                                                                                                                                & \multicolumn{1}{c}{\textbf{Synthesis}}                                                                                                                                                                   \\ 
			\\
					$
					\Theta = \left\{ 
					\mtrx{W}\ |
					\begin{array}{l}
					\mtrx{W}_{\Omega_r} = \mtrx{Y}_{\Omega_r};\\
					 \mtrx{W}_{\Omega_+} \succcurlyeq \mtrx{Y}_{\Omega_+}; \\ 
					 \mtrx{W}_{\Omega_-} \preccurlyeq \mtrx{Y}_{\Omega_-}.\end{array}\right\}
					$
					& 
					$
					\Theta = \left\{ 
					\mtrx{W}\ |
					\begin{array}{l}
					(\mtrx{D}\mtrx{W})_{\Omega_r} = \mtrx{Y}_{\Omega_r};\\
					(\mtrx{D}\mtrx{W})_{\Omega_+} \succcurlyeq \mtrx{Y}_{\Omega_+}; \\ 
					(\mtrx{D}\mtrx{W})_{\Omega_-} \preccurlyeq \mtrx{Y}_{\Omega_-}.\end{array}\right\}
					$
					\\
					\\
					$\mtrx{M} = \mtrx{A} \Cset{P}{L}, \ing{P} \geq \ing{L}$                                                                                                                                                                                                                   &$\mtrx{M} = \mtrx{I} \Cset{L}{L}$,                                                                                                                                                            \\
					$\mathcal{S}_{\mu} (\cdot)= \mathcal{H}_{\ing{P} - \mu} (\cdot)$                                                                                                                                                                &$\mathcal{S}_{\mu} (\cdot)= \mathcal{H}_{\ing{S} - \mu} (\cdot)$,                                                                                                                                        \\
					$\iter{\mu}{0} = \ing{P}-1$
					&
					$\iter{\mu}{0} = \ing{S}-1$
					\\
					$F : \mu \mapsto \mu - 1$
					&
					$F : \mu \mapsto \mu - 1$
					\\
					$\iter{\mtrx{z}}{0} = \mtrx{A}\mtrx{y}$
					&
					$\iter{\mtrx{z}}{0} = \htransp{\mtrx{D}}\mtrx{y}$
				\end{tabular}%
\end{table}

Iterating Algorithm \ref{alg:AbstractAlgo} with the parameters described above gives a declipped estimate $\hat{\mtrx{w}}$ such that:
\[
\hat{\mtrx{w}} := 
\mathcal{G}(
\Theta, \mtrx{M}, \{\mathcal{S}_{\mu}(\cdot)\}_{\mu}, \iter{\mu}{\ing{0}}, F, \iter{\mtrx{Z}}{0}, \beta, \ing{i}_{\max}
).
\]
We recall that for the analysis version $\hat{\mtrx{x}} := \hat{\mtrx{w}}$, while for the synthesis version $\hat{\mtrx{x}} := \mtrx{D}\hat{\mtrx{w}}$.

\subsubsection{Social sparse audio declippers}\label{sec:SocSparseDeclippers}

Similarly to the social sparse audio denoising procedure, we change the sparsifying operator to $\mathcal{S}_{\mu}^{\text{PEW}}(\cdot\mid\Gamma)$ and the update rule which we set now to $F_{\alpha}: \mu \mapsto \alpha \mu$. The initial value $\iter{\mu}{0}$ may also depend here on the pattern $\Gamma$ and will be precised in Section~\ref{sec:expe_declipping}.

The resulting parameters are summarized in Table~\ref{tab:SocialSparseDeclip}.

\begin{table}[h!]
	\caption{Parameters of Algorithm~\ref{alg:AbstractAlgo} for the Social Sparse Declipper \label{tab:SocialSparseDeclip}}
	\centering
		\begin{tabular}{m{0.45\columnwidth}|m{0.55\columnwidth}}
			\multicolumn{1}{c|}{\textbf{Analysis}}                                                                                                                                                                & \multicolumn{1}{c}{\textbf{Synthesis}}                                                                                                                                                                   \\ 
			\\
			$
			\Theta = \left\{ 
			\mtrx{W}\ |
			\begin{array}{l l}
			\mtrx{W}_{\Omega_r} = \mtrx{Y}_{\Omega_r};\\
			\mtrx{W}_{\Omega_+} \succcurlyeq \mtrx{Y}_{\Omega_+}; \\ 
			 \mtrx{W}_{\Omega_-} \preccurlyeq \mtrx{Y}_{\Omega_-}.\end{array}\right\}
			$
			& 
			$
			\Theta = \left\{ 
			\mtrx{W}\ |
			\begin{array}{l l}
			(\mtrx{D}\mtrx{W})_{\Omega_r} = \mtrx{Y}_{\Omega_r};\\
			(\mtrx{D}\mtrx{W})_{\Omega_+} \succcurlyeq \mtrx{Y}_{\Omega_+}; \\ 
			(\mtrx{D}\mtrx{W})_{\Omega_-} \preccurlyeq \mtrx{Y}_{\Omega_-}.\end{array}\right\}
			$
			\\
			\\
			$\mtrx{M} = \mtrx{A} \Cset{P}{L}, \ing{P} \geq \ing{L}$                                                                                                                                                                                                                   & $\mtrx{M} = \mtrx{I} \Cset{L}{L}$,                                                                                                                                                            \\
			$\mathcal{S}_{\mu} (\cdot)= \mathcal{S}^{\text{PEW}}_{\mu}(\cdot|\Gamma)$                                                                                                                                                                & $\mathcal{S}_{\mu} (\cdot)= \mathcal{S}^{\text{PEW}}_{\mu}(\cdot|\Gamma)$,                                                                                                                                        \\
			$\iter{\mu}{0}$: see Section~\ref{sec:expe_declipping}
			&
			$\iter{\mu}{0}$: see Section~\ref{sec:expe_declipping}
			\\
			$F = F_{\alpha}: \mu \mapsto \alpha \mu$
			&
			$F = F_{\alpha}: \mu \mapsto \alpha \mu$
			\\
			$\iter{\mtrx{z}}{0} = \mtrx{A}\mtrx{y}$
			&
			$\iter{\mtrx{z}}{0} = \htransp{\mtrx{D}}\mtrx{y}$
		\end{tabular}%
\end{table}

The social declipper with a \emph{predefined} time-frequency pattern $\Gamma$ is compactly written as:

\begin{eqnarray*}
	\begin{bmatrix}
		\hat{\mtrx{w}}(\Gamma)\\ 
		\mu(\Gamma)\\ 
		\mtrx{Z}(\Gamma)
	\end{bmatrix} := \mathcal{G}(
	\Theta, \mtrx{M}, \{\mathcal{S}^{\text{PEW}}_{\mu}(\cdot|\Gamma)\}_{\mu}, \iter{\mu}{\ing{0}}, F_{\alpha}, \iter{\mtrx{Z}}{0}, \beta, \ing{i}_{\max}
	),
\end{eqnarray*}

\noindent The adaptive social declipper uses this to select the ``optimal'' pattern $\Gamma$ within a prescribed collection $\left\{ \Gamma_{\ing{k}}\right\}_{\ing{k}=1}^{\ing{K}}$ for the processed signal region, by running few iterations of the algorithm (typically $\ing{i}_{\max}^{\texttt{small}} = 10$). The whiteness of the residual is evaluated with the same entropy criterion~\eqref{eq:entropy} as in denoising, which maximization yields the selected pattern $\Gamma_{\ing{k}^{\star}}$.

Correspondingly, the first value $\iter{\mu_{(\ing{k})}}{0}$ and the update rule $F_{\alpha}$ as well as the time-frequency patterns $\left\{ \Gamma_{\ing{k}}\right\}_{\ing{k}=1}^{\ing{K}}$ are essential for the algorithm to provide improvements. These will be specified in Section \ref{sec:expe_declipping}.

Once the best time-frequency pattern is selected, we run Algorithm~\ref{alg:AbstractAlgo} with the parameters listed in Table~\ref{tab:SocialSparseDeclip} and a sufficiently large $\ing{i}_{\max}$ (typically $\ing{i}_{\max}^{\texttt{large}} =10^6$) to get
\[
\hat{\mtrx{w}}:=\mathcal{G}(
\Theta, \mtrx{M}, \{\mathcal{S}^{\text{PEW}}_{\mu}(\cdot|\Gamma_{\ing{k}^{\star}})\}_{\mu}, 
\mu_{\ing{k}^{\star}}, F_{\alpha}, \mtrx{Z}_{\ing{k}^{\star}},  \beta, \ing{i}_{\max}^{\texttt{large}}
).
\] 
The pseudo-code of the adaptive social declipper for a given block of adjacent frames $\mtrx{Y} \Rset{\ing{L}}{\ing{(2b+1)}}$ is given in Algorithm \ref{alg:AUDASCITYDeclip}. Again, for the analysis version $\hat{\mtrx{x}} := \hat{\mtrx{w}}$, while for the synthesis version $\hat{\mtrx{x}} := \mtrx{D}\hat{\mtrx{w}}$.

\begin{algorithm}
	\caption{Adaptive Social Sparse Declipper 	\label{alg:AUDASCITYDeclip}}
	
	\begin{algorithmic} 
		\REQUIRE $\mtrx{Y}$, $\varepsilon$, $\mtrx{A}$ or $\mtrx{D}$, $\left\{\Gamma_{\ing{k}}\right\}_{\ing{k}}$, $\{\iter{\mu_{\ing{k}}}{0}\}_{\ing{k}}$, $\alpha,\beta$, $\ing{i}_{\max}^{\texttt{small}}$, $\ing{i}_{\max}^{\texttt{large}}$
		\STATE set parameters from Table~\ref{tab:SocialSparseDenoise}, $\alpha=1$
		\FORALL{$\ing{k}$}
		\STATE
		\STATE \resizebox{0.45\textwidth}{!}{$\begin{bmatrix}
			\hat{\mtrx{w}}_{\ing{k}}\\ 
			\mu_{\ing{k}}\\ 
			\mtrx{Z}_{\ing{k}}
			\end{bmatrix}:= 
			\mathcal{G}(\Theta, \mtrx{M},\{\mathcal{S}_{\mu}^{\text{PEW}}(\cdot|\Gamma_{\ing{k}})\}_{\mu}, \iter{\mu_{\ing{k}}}{\ing{0}}, F_{\alpha}, \iter{\mtrx{z}}{0}, \beta,\ing{i}_{\max}^{\texttt{small}})$}
		\STATE
		\STATE Compute $\var{e}_{\ing{k}}$ as in \eqref{eq:entropy}
		\ENDFOR
		\STATE $\ing{k}^{\star} := \argmax_{\ing{k}} \var{e}_{\ing{k}}$, $\alpha=0.99$
		\STATE $\hat{\mtrx{w}}:=\mathcal{G}(
		\Theta, \mtrx{M}, \{\mathcal{S}^{\text{PEW}}_{\mu}(\cdot|\Gamma_{\ing{k}^{\star}})\}_{\mu}, 
		\mu_{\ing{k}^{\star}}, F_{\alpha}, \mtrx{Z}_{\ing{k}^{\star}},  \beta, \ing{i}_{\max}^{\texttt{large}}
		).$
		\RETURN $\hat{\mtrx{w}}$
	\end{algorithmic}
	
\end{algorithm}

\subsubsection{Overlap-add synthesis}
As in denoising, the overall declipped signal is obtained by overlap-add, here without any Wiener filtering post-processing.

\subsection{Experimental Study\label{sec:expe_declipping}}

This experimental validation aims at comparing the impact of the different modelings and saturation levels on the declipping performance for audio signals. 

\paragraph{Datasets}We direct experiments on the same material (RWC and TIMIT databases) used in the denoising section (Section~\ref{sec:expe_denoising}). Each excerpt was first amplitude normalized ($\norm{ \text{vec}(\mtrx{x})}{\infty} = 1$) then artificially clipped following the hard-clipping model in~\eqref{eq:ClipModel} for various values of $0<\tau < 1$.\\

\begin{figure*}[t]
	\centering
	\subfloat[RWC Jazz]{\includegraphics[width=5cm]{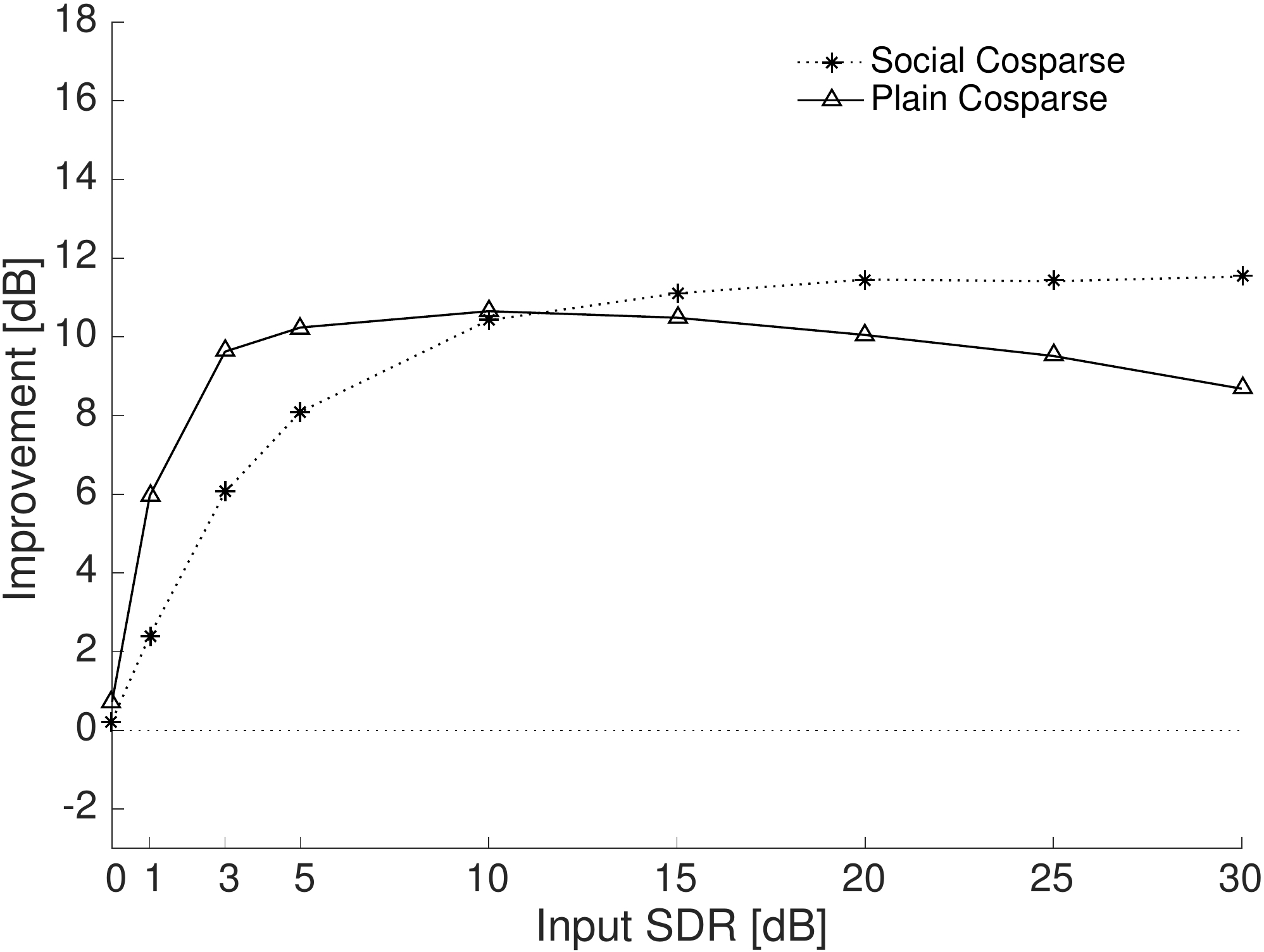}%
		\label{fig:JazzDec}}
	\hfil
	\subfloat[RWC Pop]{\includegraphics[width=5cm]{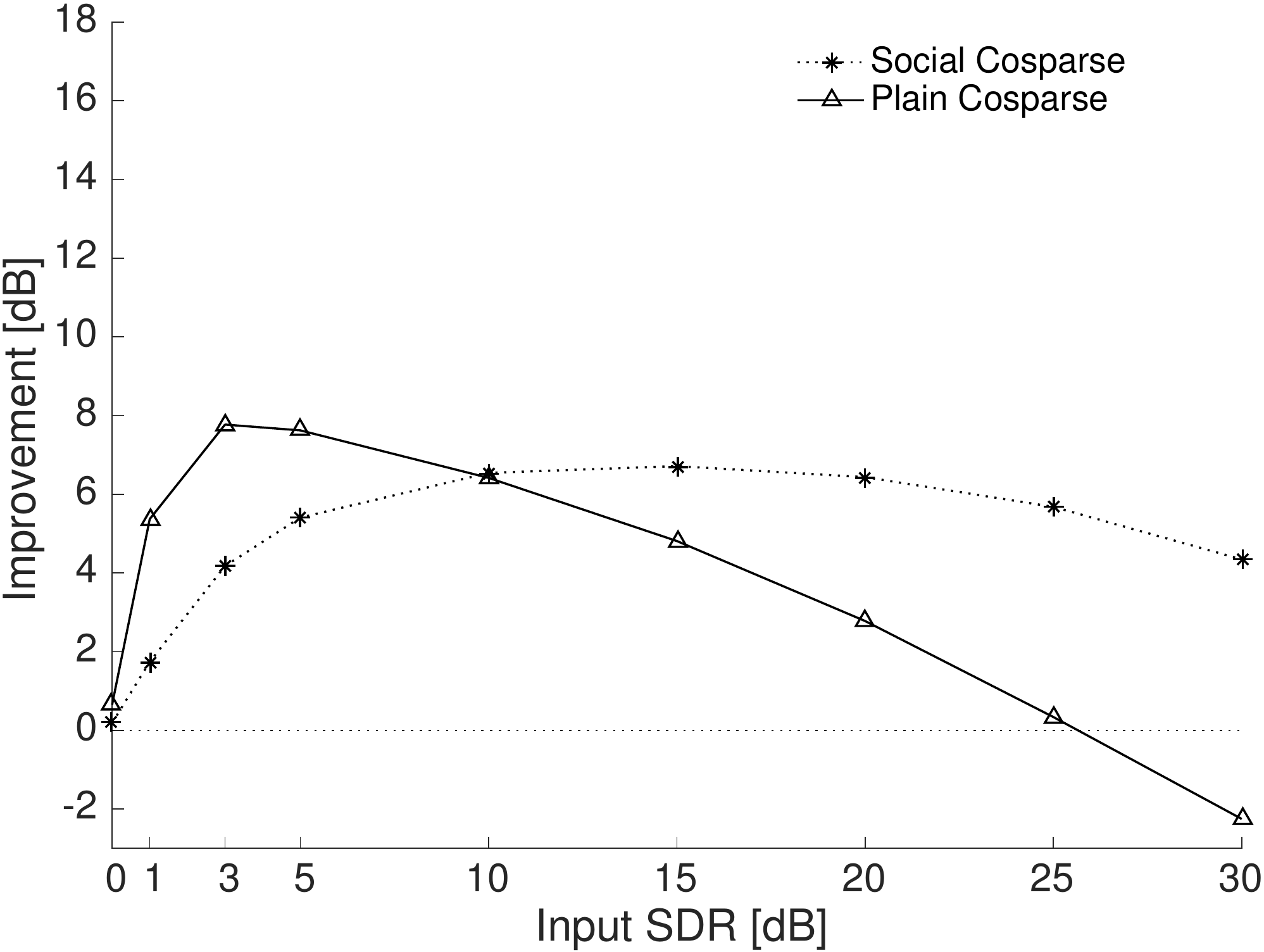}%
		\label{fig:PopDec}}
	\hfil
	\subfloat[RWC Classic: Chamber]{\includegraphics[width=5cm]{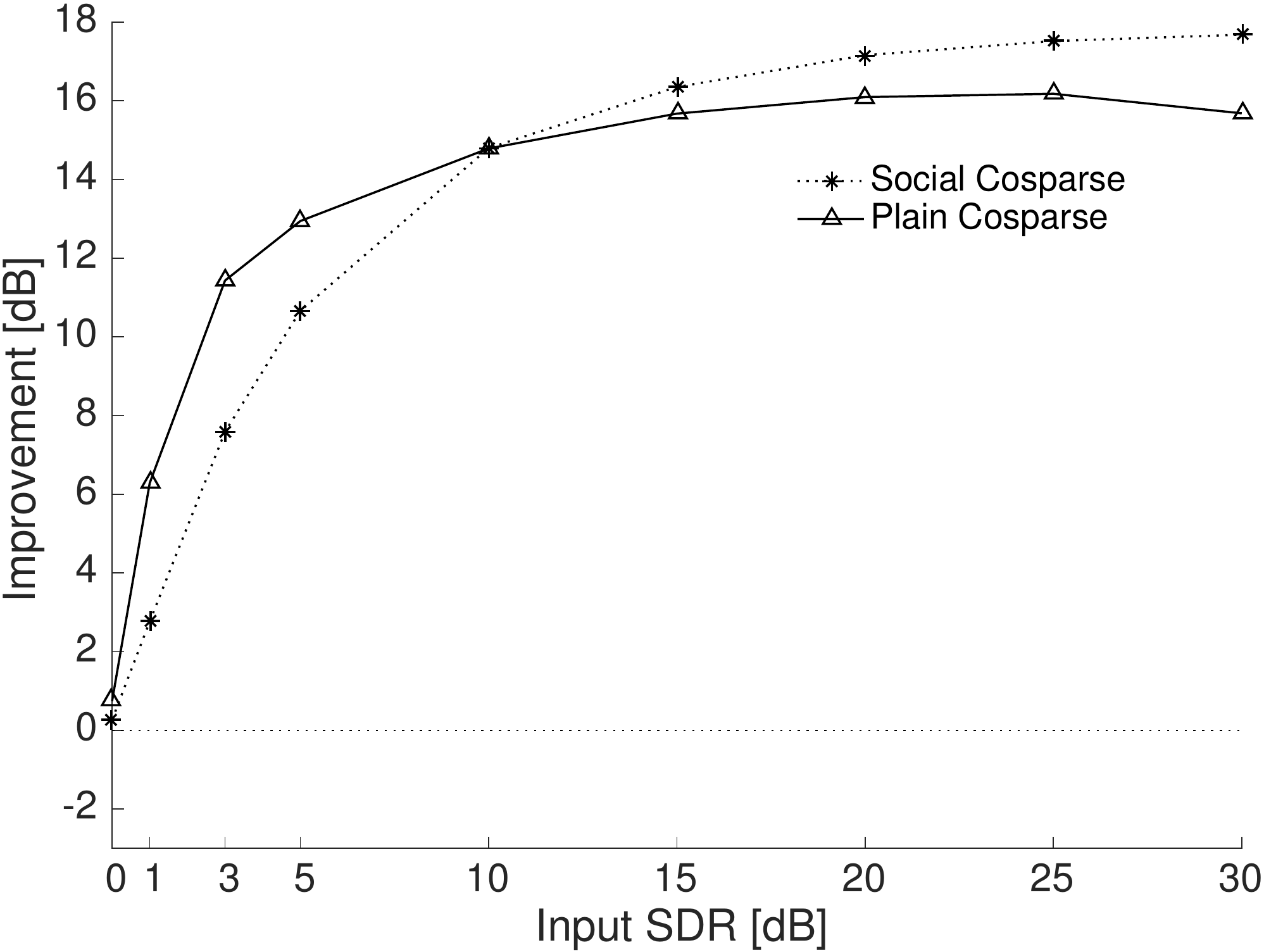}%
		\label{fig:ChamberDec}}
	\hfil
	\subfloat[RWC Classic: Vocals]{\includegraphics[width=5cm]{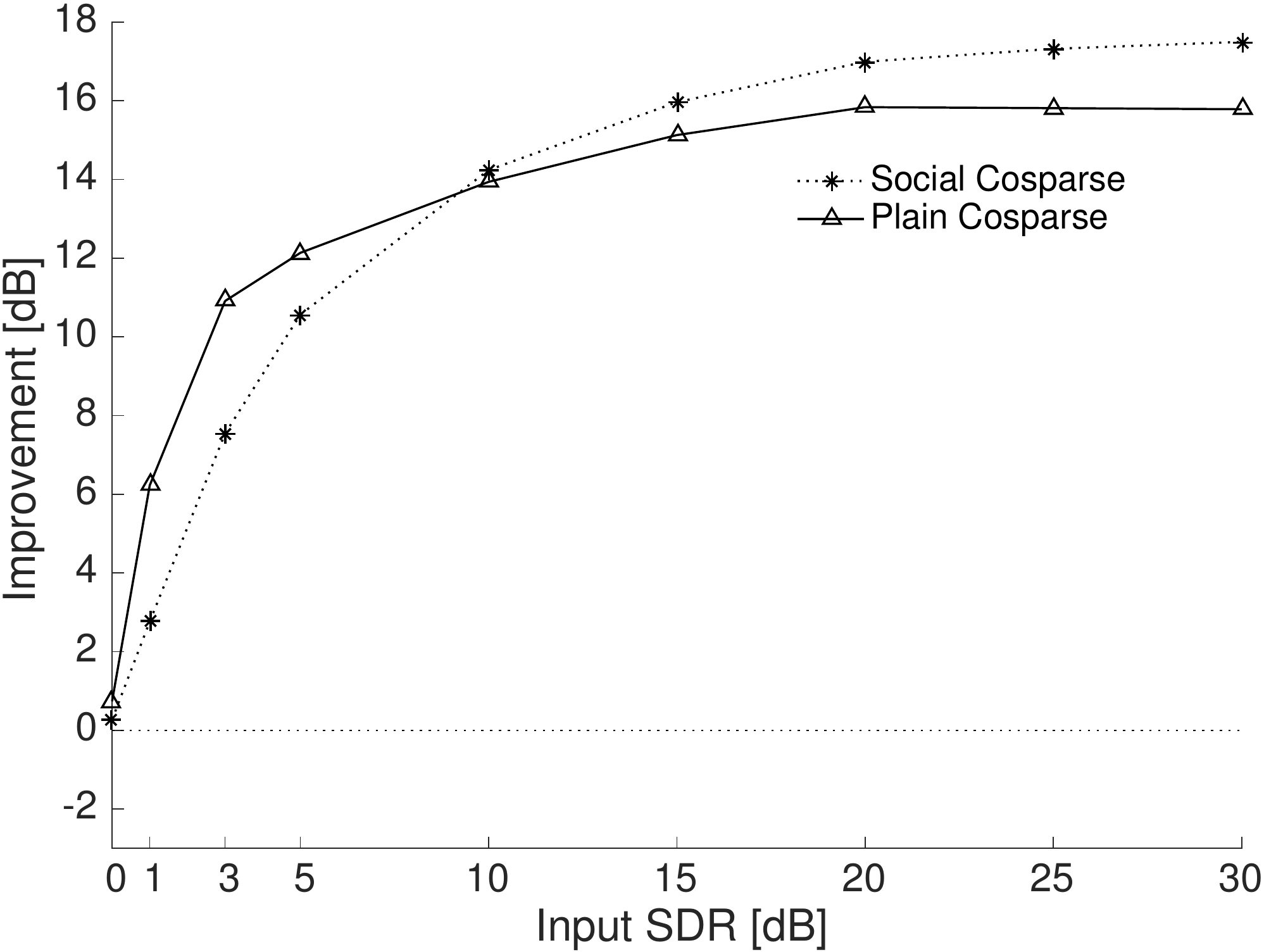}%
		\label{fig:VocalsDec}}
	\hfil
	\subfloat[RWC Classic: Symphonies]{\includegraphics[width=5cm]{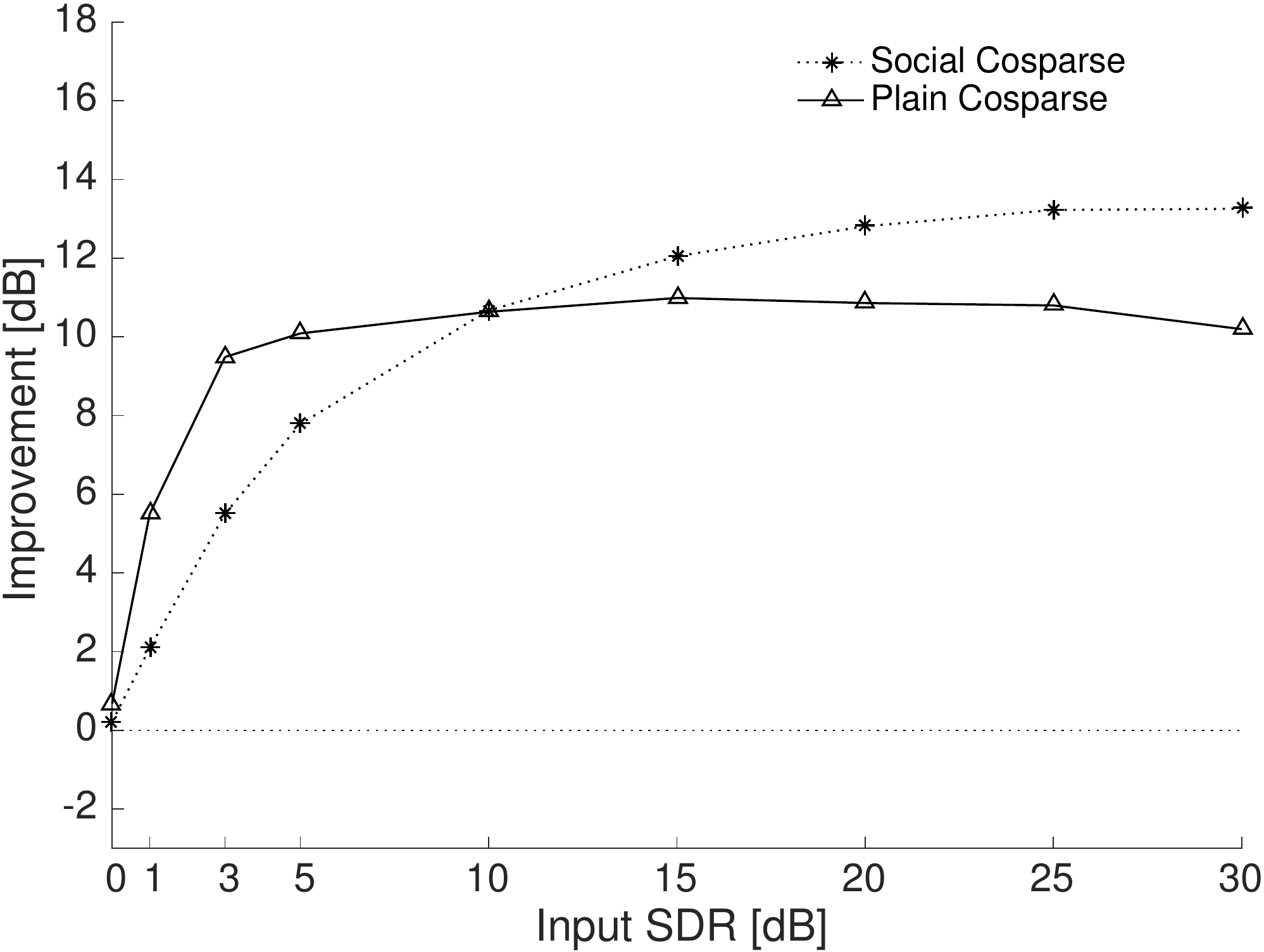}%
		\label{fig:SymphoniesDec}}
	\hfil
	\subfloat[TIMIT: Speech]{\includegraphics[width=5cm]{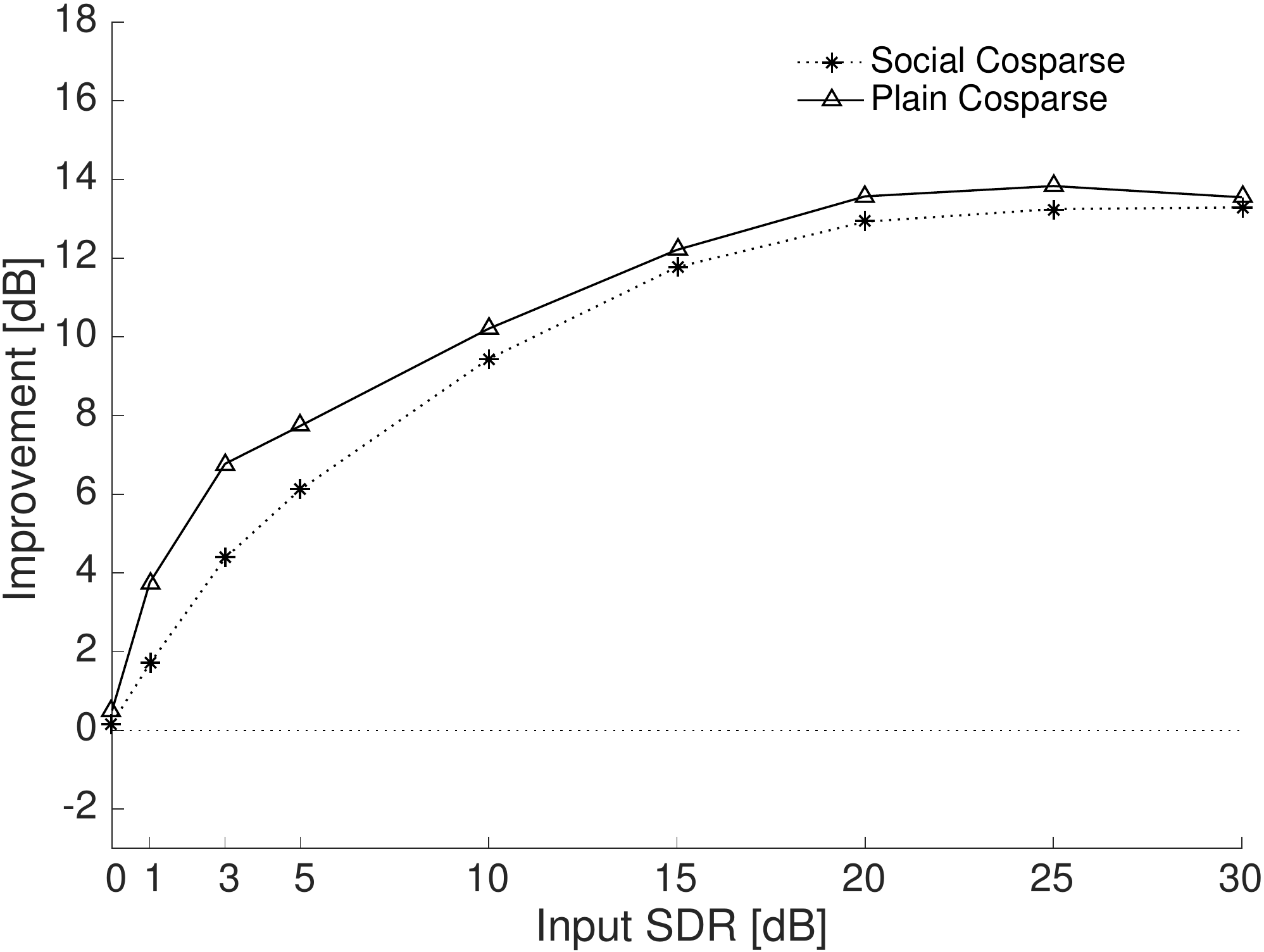}%
		\label{fig:SpeechDec}}
	\hfil	
	\caption{Numerical Results for the declipping task: SDR improvement {[}dB{]}}
	\label{fig:ResNumClip}
\end{figure*}

\begin{figure*}[ht]
	\centering
	\subfloat[TIMIT: STOI Index]{\includegraphics[width=5cm]{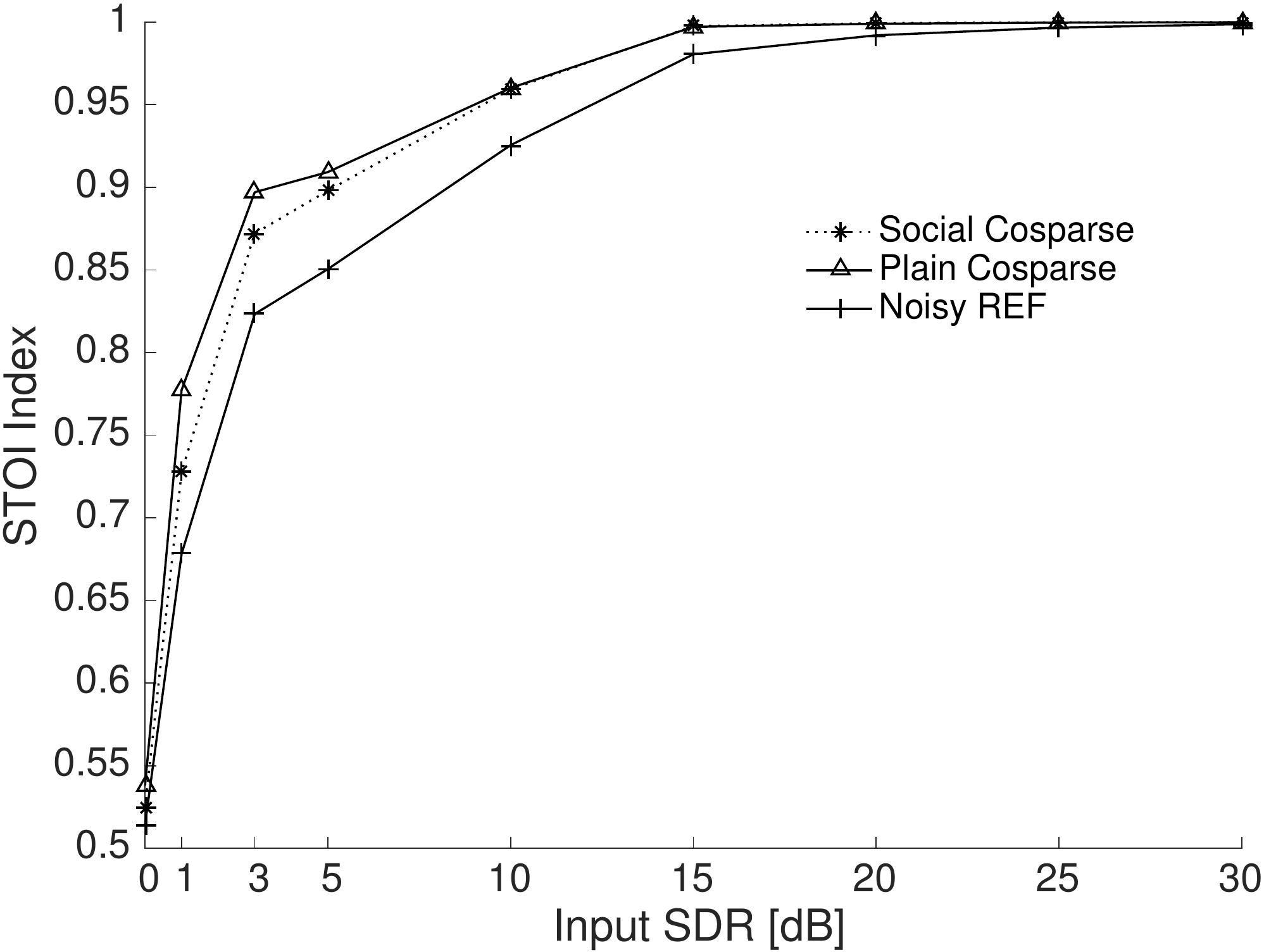}%
		\label{fig:STOITimitDec}}
	\hfil
	\subfloat[TIMIT: MOS-Mapped PESQ Value]{\includegraphics[width=5cm]{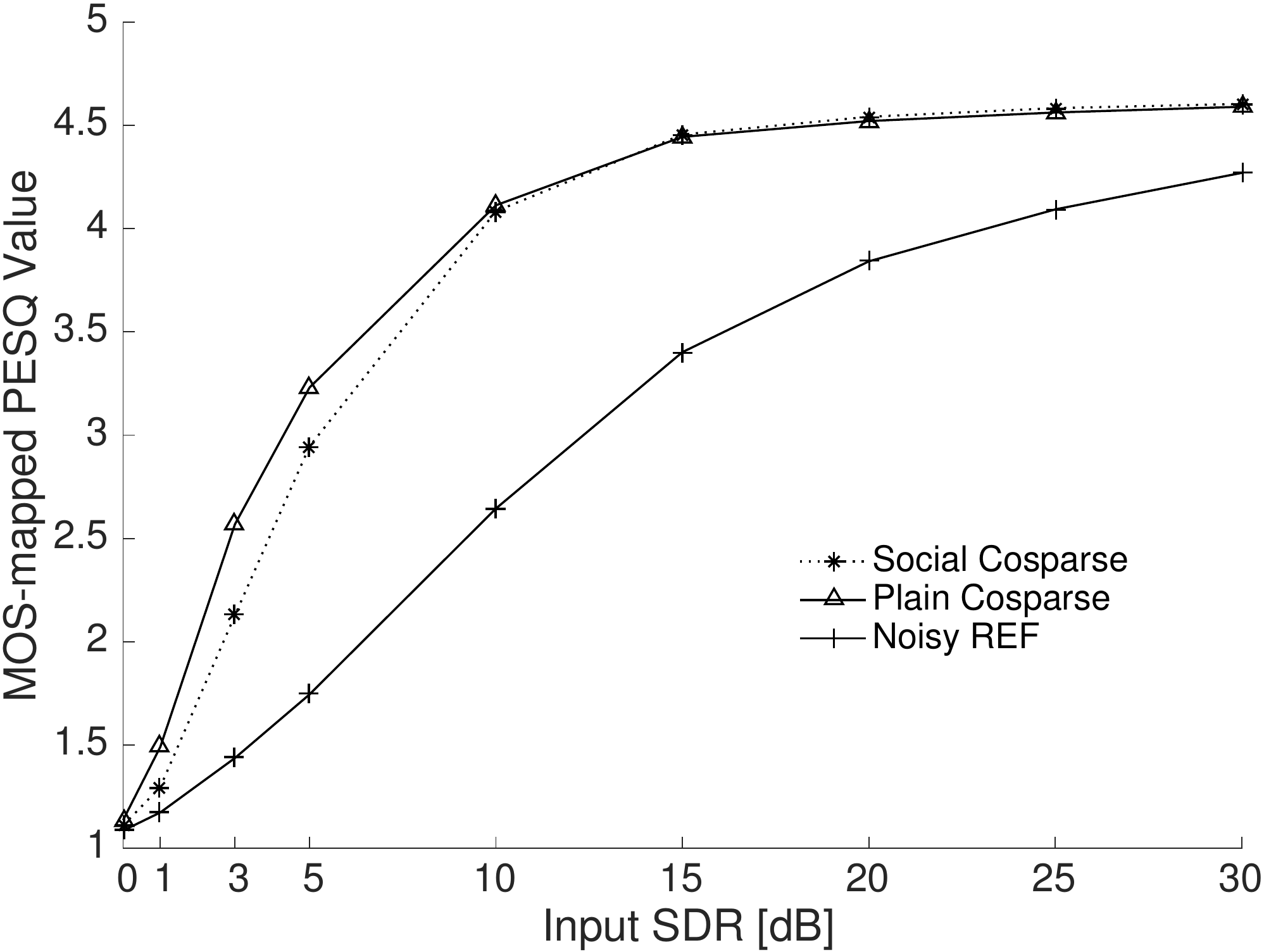}%
		\label{fig:PESQTimitDec}}
	\hfil
	\subfloat[RWC: PEAQ ODG score]{\includegraphics[width=5cm]{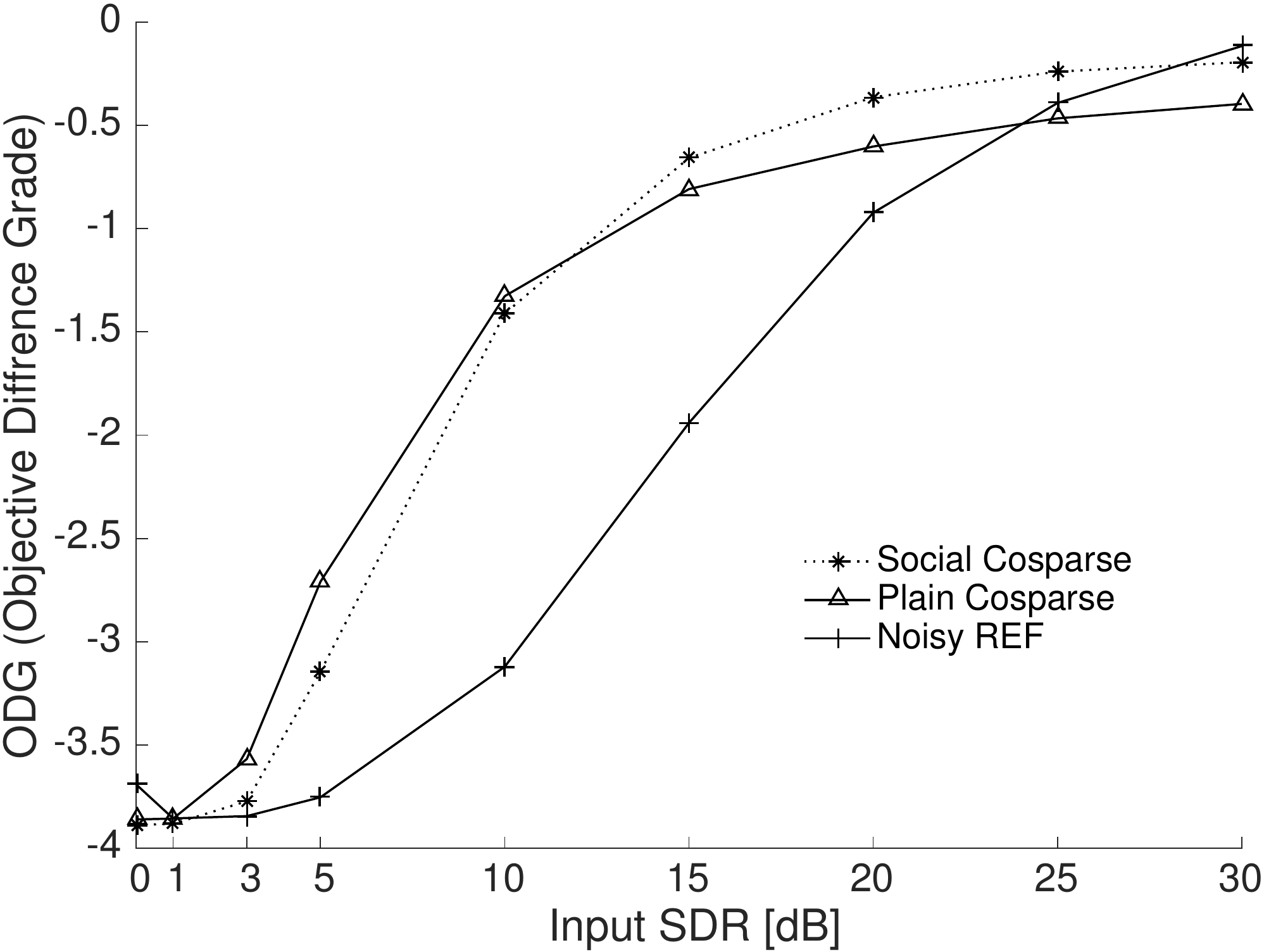}}
	\label{fig:PEAQDecRWC}
	\hfil
	\caption{Declipping Objective Quality and Intelligibility Results}
	\label{fig:ResObjSpeechClip}
\end{figure*}

\paragraph{Performance measure}We use as a first enhancement measurement the Signal-to-Distortion Ratio (SDR) \cite{vincent06:bassperf} difference ($\Delta$SDR) between the clipped and the recovered signal. For speech, we also study the perceptual quality improvements through the MOS-PESQ scores and the objective intelligibility with the STOI index. Over all musical excerpts from the RWC subset, we evaluate the perceptual audio quality with the ODG PEAQ score. Results produced on Figures~\ref{fig:ResNumClip} and~\ref{fig:ResObjSpeechClip} display averaged measurements over all available samples.

\paragraph{Compared methods}Similarly to the denoising section, we consider the plain sparse, plain cosparse, social sparse and social cosparse declippers. We set the common parameters for the algorithms as listed below.
\begin{itemize}
	\item Frame size $\ing{L} = 64$ ms for music $\ing{L} = 32$ ms for speech;
	\item Overlap, $ 75\%;$
	\item Accurracy $ \beta = 10^{-3}, \ing{i}_{\max}^{\texttt{small}}=10$, $\ing{i}_{\max}^{\texttt{large}}=10^{6};$
	\item Analysis operator, $\mtrx{A} = \text{DFT};$
	\item Synthesis operator, $\mtrx{D} = \text{inverse DFT}.$
\end{itemize}

Considering the adaptive social sparse declipper and similarly to denoising, we set the collection of time-frequency patterns $\left\{ \Gamma_{\ing{k}}\right\}_{\ing{k}=1}^{\ing{K}}$ to match the one presented on Fig. \ref{fig:Stencils} for music and Fig. \ref{fig:SpeechStencils} for speech. The specific choice of $\iter{\mu_{\ing{k}}}{0} := \norm{\Gamma}{0}\times\left(1-\tau\right)$ is motivated by the sparsity degree of the time-frequency neighborhood considered. With this parameterization, the regularization behavior is initialized inversely proportional to the maximum magnitude of the clipped signal, allowing highly clipped configurations to retain sparser regularization. Contrarily to the social sparse denoising method, we notice better improvements when the $\mu$ parameter is {\em not} updated during the initialization loop (\ie $\alpha=1$). Once the proper $\Gamma_{\ing{k}^{\star}}$ is selected, so far, we obtained the better declipping results with $\mu$ following a geometric progression of common ratio $\alpha$ with $\alpha = 0.99$. We finally set the number of overlapping segments to $\ing{b} = 5$ for music, $\ing{b} = 1$ for speech (\ie 11 frames or 3 frames at a time). \\
\paragraph{Choice of redundancy}
\begin{table}[ht]
	\centering
	\caption{Processing times comparison (plain sparse declippers)\label{tab:CompTimeDec}}
	\resizebox{0.48\textwidth}{!}{\begin{tabular}{l|c|c|c|c|c|c|c|c|}
		\cline{2-9}
		& \multicolumn{4}{c|}{Input SDR: 5 dB} & \multicolumn{4}{c|}{Input SDR: 20 dB} \\ \cline{2-9} 
		& \multicolumn{2}{c|}{Analysis} & \multicolumn{2}{c|}{Synthesis} & \multicolumn{2}{c|}{Analysis} & \multicolumn{2}{c|}{Synthesis} \\ \cline{2-9} 
		& $\Delta$SDR & $\times$RT & $\Delta$SDR & $\times$RT & $\Delta$SDR & $\times$RT & $\Delta$SDR & $\times$RT \\ \hline
		\multicolumn{1}{|l|}{$\ing{R}=1$} & 8.90 & 10.7 & 8.90 & 1183.8  & 10.37 & 7.5  & 10.37 & 1041.6 \\ \hline
		\multicolumn{1}{|l|}{$\ing{R}=2$} & 9.73 & 41.7  & 9.81 & 2353.2 & 11.04 & 24.4 & 11.13 & 1339.6  \\ \hline
		\multicolumn{1}{|l|}{$\ing{R}=4$} & 9.31 & 164.4 & 9.98 & 3782.0  & 10.55 & 85.5 & 11.21 & 2047.0  \\ \hline
	\end{tabular}}
\end{table}

As in denoising experiments, processing times in Table~\ref{tab:CompTimeDec} lead us to retain only the twice redundant setting ($\ing{R}=2$) for further comparisons.\\
\paragraph{Analysis vs synthesis}
Even if we introduce both the analysis and synthesis versions of the declippers in sections~\ref{sec:SparseDeclippers} and~\ref{sec:SocSparseDeclippers}, the detailed experimental results below will compare only the analysis social and analysis plain sparse declipping methods: while both flavors perform similarly qualitatively speaking, the computational cost of the synthesis declippers (See. Table~\ref{tab:CompTimeDec} and Section~\ref{subsec:AlgoDeclip}) is much higher. Indeed the analysis version allows a computational speedup of the order of 23 to 139 depending on the redundancy and the input SDR.\\
\paragraph{Computation time\label{par:CompTimeDec}}
\begin{table}[ht]
	\centering
	\caption{Computational performance of cosparse declippers \label{tab:DecComp}}
		\begin{tabular}{|l||c|c|c|c|}
		\hline
		\multirow{2}{*}{\begin{tabular}[c]{@{}l@{}}Input\\ SDR {[}dB{]}\end{tabular}} & \multicolumn{2}{c|}{Plain Cosparse}                                                                                              & \multicolumn{2}{c|}{Social Cosparse}                                                                                       \\ \cline{2-5} 
		&  \begin{tabular}[c]{@{}c@{}}$\Delta$ SDR\end{tabular} & \begin{tabular}[c]{@{}c@{}}$\times$ RT\end{tabular} & \begin{tabular}[c]{@{}c@{}}$\Delta$ SDR\end{tabular} & \begin{tabular}[c]{@{}c@{}}$\times$ RT\end{tabular} \\ \hline
		0 & 0.64 & 27.4 & 0.20  & 188.1 \\ \hline
		5   & 9.73  & 41.7  & 7.37  & 130.4 \\ \hline
		10   & 10.55  & 46.4  & 9.70  & 113.0 \\ \hline
		15   & 10.81  & 37.9  & 11.0  & 84.3 \\ \hline
		20   & 11.04 & 24.4  & 12.31  & 50.2 \\ \hline
	\end{tabular}
\end{table}
Table~\ref{tab:DecComp} provides processing times for both analysis social sparse (Social Cosparse) and analysis plain sparse (Plain Cosparse) declipping methods. These experiments are conducted using the same hardware and software settings described in the denoising experimental section. The cosparse declippers benefit from the same computational assets as the denoisers. We notice again that the social cosparse declipper runtime performance seems to improve as the degradation level decreases.\\

\paragraph{Comparison of declipping performance}

To evaluate the different sparse modelings, we use the shrinkages (Hard Thresholding and PEW). We also set the clipping level of the saturated material to account for different degrees of degradation. For that, we consider nine input SDR levels in dB: \{$0$, $1$, $3$, $5$, $10$, $15$, $20$, $25$, $30$\}. Results presented on Figures \ref{fig:ResNumClip} and \ref{fig:ResObjSpeechClip} show averaged measurements over all available excerpts.\\

Figure~\ref{fig:ResNumClip} shows the behavior of the two methods as a function of the input degradation level. 
For all the considered datasets, both declipping methods provide significant SDR improvements (often more than $8$dB) at (almost) all considered input SDRs. Unlike for denoising, this remains the case even for relatively high input SDRs, with one exception: the Pop category, for which the Plain Cosparse brings some degradation at very high input SDR, and the overall improvement never exceeds $8$dB. This may be due to the fact that most of the 100 unclipped excerpts in this category are mixes containing one or more tracks of dynamically compressed drums, and that at least 21 of them contain saturated guitar sounds. 

The benefit of social modeling is clear for moderate to high input SDR ($>10\text{dB}$, mild clipping), and vice-versa there is also a distinct superiority of the plain cosparse method for low input SDRs (strong clipping). Actually, the simple cosparse approach performs 2 to 4 dB better than the adaptive social method for input SDRs ranging from 1 to 5 dB on audio content from the RWC database. On the opposite, the trend tends to reverse above 10 dB input SDR as the social methods features improvements between 1 and 4 dB (even 7dB for the Pop category) above the plain cosparse technique. For speech content, the difference is less obvious yet Fig.~\ref{fig:SpeechDec}, Fig.~\ref{fig:STOITimitDec} and~\ref{fig:PESQTimitDec} displays better improvements either in terms of SDR, objective intelligibility or quality for the plain cosparse declipper.\\

Contrarily to denoising settings, standard deviation results indicate that the social cosparse declipper produces more consistent results as the standard deviation is the lowest for this technique in 67\% of the tested cases. We also observed that, 
for any of the considered algorithms, the improvement variability seems to increase with the input SDR. 
\\

The difference in performance between the plain and social cosparse declippers on music at low input SDR might come from the nature of the degradation. Indeed, contrarily to additive noise, the magnitude saturation adds broadband stripes in the time-frequency plane due to discontinuities in the time domain. 
This way, the signal's underlying structure (embodied by a time-frequency pattern $\Gamma$) is not only hidden as in the additive noise case, but also possibly distorted: during the initialization loop of the social approach, it is possible that a ``wrong'' pattern $\Gamma^{*}$ is selected. In contrast, the plain cosparse declipper cannot be affected by this type of behaviour. Another interesting result which could supports this hypothesis is that for higher SDR, the social method is actually benefiting from the time-frequency structure identification as it performs better.

\section{Conclusion}\label{sec:Conclusion}

The algorithmic framework for audio restoration proposed in this paper is versatile from many points of view. 

First, it can handle several types of audio distortion models, as demonstrated here on two audio restoration problems, denoising and declipping. More restoration scenarios are envisioned, and further work will include in particular multichannel scenarios and extensions to non-Gaussian noise. 

Second, it handles transparently both the analysis and synthesis flavors of time-frequency models.  On pilot studies, the analysis and synthesis flavors yields almost identical SNR performance on denoising and declipping, the analysis version being between 20\% and nearly 40\% faster than the synthesis one on denoising, and more than 20 times faster on declipping. In fact, real-time was achieved on experiments not shown here. The plain and social flavors have comparable qualitative restoration performance but reach their maximum speed in complementary input degradation regimes.

In terms of quality, our detailed study of the analysis version for declipping shows SDR improvements consistently exceeding $8$dB for various types of speech and music and a wide range of clipping levels. The only notable exception is the Pop dataset, possibly due to the presence of dynamically compressed drums and saturated guitar sounds. 
Similarly, in the denoising scenario, consistent SNR improvements are observed that are either on par with or better than what the widely used Block Thresholding reference algorithm can achieve, especially for input SNRs above $10$dB.  Similar trends are observed with perceptually-aware objective quality measures, however further work would be needed to confirm them on subjective listening tests.

Finally, the framework can seemlessly exploit plain or social sparsity through the integration of appropriate shrinkage operators.
For declipping, social cosparsity brings a clear benefit for moderate but still significant saturation levels (input SDR over $10$dB), while plain cosparsity seems consistently preferable for strongly clipped scenarios. This is possibly due to the difficulty to properly detect social sparsity patterns heavily corrupted by strong saturation, and calls for further studies to guide the choice of such patterns possibly based on learning techniques. 
For denoising, the behaviour is opposite, as the  performance of the plain and social flavors of the framework seem relatively similar except at low input SNR where social sparsity brings a mild benefit compared to the plain one. 

\appendices
\section{Generalized projection for denoising}\label{app:ProxDenoising}
The goal is to solve~\eqref{eq:ConstrainedProjection}, i.e.,
\[
 	\minim_{\mtrx{W} \in \Theta} \norm{\mtrx{M} \mtrx{W} - \mtrx{Z}}{\text{F}}. 
\]
with $\mtrx{M} = \mtrx{A}$,  $\Theta = \{\mtrx{W}: \norm{\mtrx{w}-\mtrx{y}}{\text{F}}\leq\varepsilon\}$ for the analysis case, and $\mtrx{M} = \mtrx{I}$ and $\Theta = \{\mtrx{W}: \norm{\mtrx{D}\mtrx{w}-\mtrx{y}}{\text{F}}\leq\varepsilon\}$ for the synthesis case.
For the synthesis case, this is more explicitly 
\[
 	\minim_{\mtrx{W}} \norm{\mtrx{W} - \mtrx{Z}}{\text{F}}^{2} 
	\subjto \norm{\mtrx{D}\mtrx{W}-\mtrx{Y}}{\text{F}}^{2} \leq \varepsilon^{2}.
\]
Let us now show that in the analysis case the optimization problem can be cast to a similar form.
Since we consider a tight frame $\htransp{\mtrx{A}} \mtrx{A}=\zeta \mtrx{I}$, the orthogonal projection onto the linear span of $\mtrx{A}$ is $P_{\mtrx{A}} = \zeta^{-1} \mtrx{A}\htransp{\mtrx{A}}$ 
and for any $\mtrx{W},\mtrx{Z}$, 
\begin{eqnarray*}
\norm{\mtrx{A}\mtrx{W}-\mtrx{Z}}{\text{F}}^{2}
&=& 
\norm{\mtrx{A}\mtrx{W}-P_{\mtrx{A}}\mtrx{Z}+(\mtrx{I}-P_{\mtrx{A}})\mtrx{Z}}{\text{F}}^{2}\\
&=& 
\norm{\mtrx{A}\mtrx{W}-P_{\mtrx{A}}\mtrx{Z}}{\text{F}}^{2}
+
\norm{(\mtrx{I}-P_{\mtrx{A}})\mtrx{Z}}{\text{F}}^{2}\\
&=& 
\norm{\mtrx{A}(\mtrx{W}-\zeta^{-1}\htransp{\mtrx{A}}\mtrx{Z})}{\text{F}}^{2}
+
\norm{(\mtrx{I}-P_{\mtrx{A}})\mtrx{Z}}{\text{F}}^{2}\\
&=& 
\zeta \norm{\mtrx{W}-\zeta^{-1}\htransp{\mtrx{A}}\mtrx{Z}}{\text{F}}^{2}
+
\norm{(\mtrx{I}-P_{\mtrx{A}})\mtrx{Z}}{\text{F}}^{2}.
\end{eqnarray*}
Minimizing the left hand side with the constraint $\mtrx{W} \in \Theta$ is thus equivalent to 
\[
 	\minim_{\mtrx{W}} \norm{\mtrx{W} - \zeta^{-1}\htransp{\mtrx{A}}\mtrx{Z}}{\text{F}}^{2} 
	\subjto \norm{\mtrx{W}-\mtrx{Y}}{\text{F}}^{2} \leq \varepsilon^{2}.
\]
Both cases boil down to an optimization problem
	
	\begin{equation}\label{eq:GeneralLsqIneq}
		\hat{\mtrx{w}} = \argmin_{\mtrx{w}} \norm{\mtrx{w}-\mtrx{B}}{\text{F}}^2 \subjto \norm{\mtrx{f}\mtrx{w}-\mtrx{y}}{\text{F}}\leq\varepsilon
	\end{equation}
	with $\mtrx{B} = \zeta^{-1} \htransp{\mtrx{A}}\mtrx{Z}$ and $\mtrx{F} = \mtrx{I}$ for the analysis case, while $\mtrx{B} = \mtrx{Z}$ and $\mtrx{F} = \mtrx{D}$ for the synthesis case. When $\mtrx{F}$ is a tight frame, $\mtrx{F}\htransp{\mtrx{F}} = \xi \mtrx{I}$,  problem~\eqref{eq:GeneralLsqIneq} has a  closed form solution \cite[Section 2]{yang2013linearized}

	\begin{equation}\label{eq:GeneralProx}
		\hat{\mtrx{w}} = \mtrx{b} 
		- \frac{1}{\xi} \left(\frac{\norm{\mtrx{f}\mtrx{b} -\mtrx{y}}{\text{F}}-\varepsilon}{\norm{\mtrx{f}\mtrx{b} -\mtrx{y}}{\text{F}}}\right)_{+}
		\cdot\htransp{\mtrx{f}}(\mtrx{f}\mtrx{b} - \mtrx{y}).
	\end{equation}

\section{Generalized projection for declipping}\label{app:ProxDeclipping}

The goal is to solve~\eqref{eq:ConstrainedProjection}, i.e.,
\[
 	\minim_{\mtrx{W} \in \Theta} \norm{\mtrx{M} \mtrx{W} - \mtrx{Z}}{\text{F}}. 
\]
with some constraint set $\Theta$. 

In the analysis case, as shown in Appendix~\ref{app:ProxDenoising}, as soon as $\htransp{\mtrx{A}}\mtrx{A} = \mtrx{I}$, minimizing $\norm{\mtrx{A}\mtrx{W}-\mtrx{Z}}{\text{F}}^{2}$ under the constraint
\[
\mtrx{W} \in \Theta := \left\{ 
\begin{array}{l l}
 &  \mtrx{W}_{(\ing{ij})} = \mtrx{Y}_{(\ing{ij})},\ \ing{ij} \in \Omega_r;\\
\mtrx{W}\mid &  \mtrx{W}_{(\ing{ij})} \geq \tau, \ing{ij} \in \Omega_+; \\ 
 &   \mtrx{W}_{(\ing{ij})} \leq -\tau,\ \ing{ij} \in \Omega_-.\end{array} \right\}
\]
 is equivalent to minimizing $\norm{\mtrx{W}-\htransp{\mtrx{A}}\mtrx{Z}}{\text{F}}^{2}$ under the constraint $\mtrx{W} \in \Theta$. 
As the contraint is written componentwise, the optimization can be done componentwise yielding
		\begin{equation}
			\hat{\mtrx{w}}_{(\ing{ij})} = \left\{ 
			\begin{array}{l l}
			\mtrx{Y}_{(\ing{ij})} & \text{if } \ing{ij}\in\Omega_r;\\
			(\htransp{\mtrx{A}}\mtrx{Z})_{(\ing{ij})} & \text{if~}\left\{\begin{array}{l}\ing{ij}\in\Omega_+,(\htransp{\mtrx{A}}\mtrx{Z})_{(\ing{ij})} \geq \tau;\\\text{or}\\\ing{ij}\in\Omega_-,(\htransp{\mtrx{A}}\mtrx{Z})_{(\ing{ij})} \leq -\tau;\end{array}\right.\\
			\sign(\mtrx{Y}_{\ing{ij}})\tau & \text{otherwise.}\\
			\end{array} \right.
		\end{equation}

For the synthesis case, $\mtrx{M}=\mtrx{I}$ and 

$$\Theta := \left\{ 
\mtrx{W}\ |
\begin{array}{l}
(\mtrx{D}\mtrx{W})_{\Omega_r} = \mtrx{Y}_{\Omega_r};\\
(\mtrx{D}\mtrx{W})_{\Omega_+} \succcurlyeq \mtrx{Y}_{\Omega_+}; \\ 
(\mtrx{D}\mtrx{W})_{\Omega_-} \preccurlyeq \mtrx{Y}_{\Omega_-}.\end{array} \right\}.$$

\noindent The corresponding optimization problem for the synthesis case writes:
	\begin{equation}\label{eq:SynthesisCliplsq}
	\hat{\mtrx{w}}=\argmin_{\mtrx{w}}\norm{\mtrx{w}-\mtrx{z}}{\text{F}}^2 \subjto \mtrx{d}\mtrx{w}\in\Theta.
	\end{equation}

	Unfortunately, as the magnitude constraint is expressed here in the frequency domain, there is no simple expression for $\hat{\mtrx{w}}$.  As explained in~\cite{kiti:tel-01237323}, we can solve~\eqref{eq:SynthesisCliplsq} with an iterative procedure, leading to the much higher cost of the synthesis (non)social sparse declipper.

\section*{Acknowledgment}

This work benefits from the financial support of the European Research Council, PLEASE project (ERC-StG-2011-277906), and of the Brittany Region (ARED 9173). The authors would like to thank Matthieu Kowalski for precious discussion and advice.




%
\bibliographystyle{IEEEtran}
\bibliography{bibliography}


%

%
%




\end{document}